\documentclass[sigconf,screen]{acmart}

\makeatletter
\def\subsubsection{\@startsection{subsubsection}{3}%
 \z@{.5\linespacing\@plus.7\linespacing}{.1\linespacing}%
 {\it\sf\bf}}
\makeatother  

\makeatletter
\renewcommand\@formatdoi[1]{\ignorespaces}
\makeatother
\renewcommand\footnotetextcopyrightpermission[1]{}

\usepackage{multirow}
\usepackage{tcolorbox}
\usepackage{booktabs} 
\usepackage{amsmath}
\usepackage{graphicx}
\usepackage{xcolor}
\usepackage{csquotes}
\usepackage{array}
\usepackage{cuted}
\usepackage[linesnumbered,ruled,vlined]{algorithm2e}
\SetCommentSty{\color{blue!70}}
\usepackage{epstopdf}

\usepackage{afterpage}
\usepackage{makecell}
\usepackage{listings}
\usepackage[normalem]{ulem}
\definecolor{codegreen}{rgb}{0,0.6,0}
\definecolor{codegray}{rgb}{0.5,0.5,0.5}
\definecolor{codepurple}{rgb}{0.58,0,0.82}
\definecolor{backcolour}{rgb}{0.95,0.95,0.92}
\newcommand{\myline}[1]{{\medskip\noindent\textbf{#1.}}}

\usepackage{url}
\usepackage{hyperref}
\usepackage{enumerate}
\usepackage[labelfont=bf]{caption}
\usepackage{xcolor}
\usepackage{color, colortbl}

\usepackage[utf8]{inputenc}
\usepackage{amsthm}
\usepackage[linesnumbered,ruled,vlined]{algorithm2e}

\SetKw{ParallelForEach}{parallel \ForEach}

\newcommand{\sys}{PostgreSQL-V~2.0}
\newcommand{\cidr}{PostgreSQL-V~1.0}

\newcommand{\ivfflat}{IVF\_FLAT}

\newcommand{\hnsw}{HNSW}

\usepackage{listings}
\usepackage{pifont}

\lstdefinestyle{mystyle}{
  backgroundcolor=\color{backcolour},  
  commentstyle=\color{codegreen},
  keywordstyle=\color{magenta},
  numberstyle=\tiny\color{codegray},
  stringstyle=\color{codepurple},
  basicstyle=\ttfamily\small,
  breakatwhitespace=false,
  breaklines=true,         
  captionpos=b,          
  keepspaces=true,         
  numbersep=5pt,         
  showspaces=false,        
  showstringspaces=false,
  showtabs=false,
  otherkeywords = {<<, >>, WITH}, 
  tabsize=2
}

\usepackage{tikz}
\usetikzlibrary{shapes.geometric}

\begin{document}
\pagestyle{plain} 

\title{Building An Integrated Vector Database System in PostgreSQL}

\author{Jiayi Liu \;\; Te Guo \;\; Jianguo Wang}
\affiliation{
\vspace{0.1cm}
 \institution{\textsf{Purdue University}}
 \vspace{0.1cm}
  \city{\textit{\{liu4127; guo777; csjgwang\}@purdue.edu}}
}

\begin{abstract}

This paper presents \sys{}, a scalable integrated vector database system inside PostgreSQL. Existing PostgreSQL-based vector search systems such as pgvector, embed vector indexes into PostgreSQL's page-oriented storage engine, incurring significant overhead that leads to a huge performance gap with specialized vector databases. In our earlier work, we introduced \cidr{}, which addresses this issue by separating vector index structures from PostgreSQL's storage engine, enabling vector search performance close to that of native vector index libraries while preserving SQL compatibility. However, we find that \cidr{} has three limitations that matter for real-world workloads: it only supports a single connection (without concurrency), recovery time grows with index size, and physical replication is unsupported.

We further present \sys{}, which closes all three gaps. \sys{}'s \textbf{concurrency support} enables fully concurrent vector searches and updates across PostgreSQL's multi-process backends, delivering up to \textbf{36.4$\times$} the throughput of \cidr{} while serving 32 concurrent clients. \sys{}'s \textbf{fast crash recovery} keeps cost independent of total index size, remaining near \textbf{20~ms} while \cidr{}'s grows into seconds-scale. \sys{}'s \textbf{physical replication support} extends physical replication to the decoupled index, preserving index consistency on standbys without burdening the primary node. Together, these advances make \sys{} a fully concurrent, crash-resilient, and replication-ready vector database inside PostgreSQL.

\end{abstract}

\maketitle

\section{Introduction}
\label{sec:intro}

Vector similarity search, which retrieves the top-$k$ nearest vectors of a query vector under a given distance metric, has become a critical primitive in modern data infrastructure for AI, powering retrieval-augmented generation (RAG), semantic search, and recommendation systems~\cite{covington2016deep,VecDBTutorial24,gao2023retrieval,PanWL24,karpukhin2020dense,VecDBPanel24,Bruch24Book,StonebrakerP24}. It must be served at low latency, under concurrent load, and with the reliability guarantees that production deployments demand.

Two architectural approaches for vector databases have emerged. \emph{Specialized} vector databases (e.g., Faiss~\cite{GithubFaiss}, Milvus~\cite{Milvus21}, Pinecone~\cite{Pinecone}) are purpose-built for vector workloads and deliver high search performance by designing their storage and execution engine around vector search from the ground up. \emph{Integrated} systems (e.g., pgvector~\cite{Pgvector}, PASE~\cite{PASE20}, AlloyDB AI~\cite{AlloyDBVec}, GaussDB-Vector~\cite{Gaussdb-Vector}, and \cidr{}~\cite{PostgreSQLV-CIDR26}) add vector search capabilities to an existing database engine and are attractive in practice because vector data can be stored and queried alongside relational attributes in a single system, under a unified transaction and consistency model, without data migration or introducing a separate operational tier.

Integrated systems, however, have struggled to match the search performance of specialized ones. The dominant reason is architectural: existing implementations embed vector indexes into the host's storage engine, inheriting constraints that are a poor fit for vector search. PostgreSQL is a representative case: its vector search extensions (e.g., pgvector~\cite{Pgvector} and PASE~\cite{PASE20}) must express vector indexes in a page-oriented format, which introduces significant overhead relative to native vector index libraries~\cite{VecDBRDBMSICDE24} and forecloses optimizations that specialized vector databases readily adopt.

In our earlier work, published at CIDR 2026~\cite{PostgreSQLV-CIDR26}, we introduced \cidr{}, a system that addresses this gap by \emph{decoupling} the vector index from PostgreSQL's storage engine, allowing it to be backed by native libraries while remaining compatible with PostgreSQL's query engine. The decoupled index adopts an LSM-style framework for efficient updates and a lightweight consistency mechanism that keeps the index crash-safe without reintroducing the page-oriented format that decoupling was meant to escape. \cidr{} demonstrated the feasibility of the approach and achieved performance on par with native libraries.
However, there are three key limitations in \cidr{}: it serves only a single connection (i.e., concurrency is unsupported), crash recovery time grows linearly with index size, and physical replication is unsupported. Section~\ref{sec:bg-decouple} details each gap.

\myline{Contributions}
Based on \cidr{}, we present \sys{} (\textbf{Sca}lable \textbf{D}ecoupled \textbf{V}ector Database), an integrated vector database within PostgreSQL that closes the above three gaps. 

\begin{enumerate}
\item \textbf{Concurrency Support} (Section~\ref{sec:concurrency}).
\sys{} converts index access from PostgreSQL's multi-process model to a multi-thread model within a single process, enabling fully concurrent vector search and update across PostgreSQL's backends while preserving pluggable use of native vector index libraries.
It further minimizes the cold-start window after a restart through an \texttt{mmap}-based mode that shifts the bulk of the segment loading cost off the critical path, dramatically reducing the time before search can resume.

\item \textbf{Fast Crash Recovery}  (Section~\ref{sec:crash_recovery}).
\sys{} guarantees crash consistency for the decoupled index, just as \cidr{} does, but at significantly lower cost that, unlike \cidr{}'s, stays constant as the index grows. It redesigns how crash-recovery state is tracked so that recovery touches only a small portion of the index that may be inconsistent upon a crash, leaving the far larger portion of the index untouched.

\item \textbf{Physical Replication} (Section~\ref{sec:replication}).
\sys{} supports physical replication of the decoupled index via a novel two-channel design: a \emph{consistency-critical channel} preserves index consistency on the standby through custom WAL records, while a \emph{performance-critical channel} propagates index binaries asynchronously to sustain search efficiency on standbys.

\end{enumerate}

\myline{Experimental Overview} 
We evaluate \sys{} with up to 32 concurrent clients (Section~\ref{sec:ev_staticsearch}), demonstrating the benefit of concurrency support: for pure vector search, it delivers up to \textbf{36.4$\times$} the throughput of \cidr{}. \sys{} also sustains up to \textbf{4.2$\times$} the throughput of pgvector on in-memory indexes and up to \textbf{19.4$\times$} the throughput of pgvectorscale on disk-based indexes.
Under a dynamic workload of concurrent inserts, deletes, and searches (Section~\ref{sec:ev_updates}), \sys{} sustains roughly \textbf{5.1$\times$} the throughput of pgvector while maintaining stable recall as pgvector's degrades over time. 
Turning to failure resilience, we show that the recovery overhead stays independent of index size, remaining near \textbf{20~ms} even as \cidr{}'s grows to over \textbf{4~seconds} at comparable scale (Section~\ref{sec:ev_crash_recovery}), and that the \texttt{mmap} cold-start mode reduces the post-restart unavailability window by nearly two orders of magnitude (Section~\ref{sec:ev_cold_start}). 
Finally, for physical replication, we demonstrate that serving standbys imposes negligible overhead on the primary while the standby preserves search quality identical to the primary's (Section~\ref{sec:ev_replication}).

Overall, we believe this work represents a significant milestone toward enabling efficient vector search in relational databases.

\myline{Open-Source}
The source code of \sys{} is available at:\\
\url{https://github.com/purduedb/PostgreSQL-V.git}.

\section{Background}

\subsection{Implementing Vector Search in PostgreSQL}
\label{sec:bg-impl}

PostgreSQL has been extended to support vector search by several extensions, including pgvector~\cite{Pgvector}, PASE~\cite{PASE20}, and pgvectorscale~\cite{pgvectorscale}. 
We describe the major components of their design.
 
\myline{Storing embeddings}
Embeddings are stored as values of a dedicated column data type (e.g., the \texttt{vector} type introduced by pgvector) that holds a fixed-dimensional vector. A table therefore keeps each vector alongside its relational attributes in the same row, so vector and non-vector data are managed under one schema.
 
\myline{Vector indexes via \texttt{IndexAmRoutine}}
A vector index is registered as a new index access method through PostgreSQL's \texttt{IndexAmRoutine} interface, the same extension point used by built-in indexes such as B-tree. The interface defines the callbacks that PostgreSQL's core engine invokes over the lifetime of an index: the build callback that constructs an index over a vector column, the insert callback that adds an entry for a new tuple (the storage format of a table row), the scan callbacks that return the matching tuple identifiers (TIDs) in order, and the vacuum callbacks that remove index entries for dead tuples. By implementing this interface, an extension exposes a vector index that the executor can use transparently. A top-$k$ similarity query is then written in plain SQL as an \texttt{ORDER BY} on a distance operator (e.g., $<\texttt{column}>$ \texttt{<->} $<\texttt{query}>$) together with a \texttt{LIMIT}~$k$ clause. 
When a vector index is available on the column, PostgreSQL answers the top-$k$ query with an index scan, invoking the scan callbacks described above to return the $k$ nearest vectors as TIDs. The executor then fetches the corresponding rows from the heap (PostgreSQL's table storage) and applies a visibility check under PostgreSQL's MVCC (multi-version concurrency control), so that only rows visible to the current transaction are returned.

\myline{Hybrid search} 
Because vectors live in ordinary tables, a query may combine an \texttt{ORDER BY}/\texttt{LIMIT} top-$k$ clause with other relational predicates (e.g., a \texttt{WHERE} filter over scalar attributes such as a category or time range), so the database can further restrict the search results within the same statement. These predicates are evaluated by PostgreSQL's executor on the heap tuples fetched after the index scan, that is, hybrid search uses post-filtering (e.g., in pgvector~\cite{Pgvector}) without any involvement of the vector index. Combining vector similarity with relational operations in a single query is a practical advantage of building vector search into a relational engine.
 
\myline{Reuse of PostgreSQL architecture}
Because these extensions store their index data in PostgreSQL's storage engine and route all accesses through its buffer manager, the vector index is managed along with the heap data pages and inherits PostgreSQL's mature infrastructure. 
In particular, a vector index update and its corresponding heap update within a transaction both emit write-ahead log (WAL) records that are force flushed at commit, so both are recovered together after a crash via log replay, keeping the index consistent with the heap. 
The same WAL mechanism also supports replication: a standby replays the WAL records, reproducing the heap's and index's modifications and staying up to date with the primary. By conforming to the page-oriented index framework, these extensions obtain durability, index consistency, and support for replication without reimplementing them.
 
\subsection{Limitations of Vector Search in PostgreSQL}
\label{sec:bg-limit}

The convenience of reusing PostgreSQL's architecture comes at a cost. These extensions must integrate the vector index into PostgreSQL's storage engine, which requires expressing it in a page-oriented format. This requirement is a poor match for vector indexes, and we highlight three resulting limitations.

\myline{Performance degradation}
Modern vector indexes (e.g., HNSW~\cite{HNSW18}, IVF\_FLAT~\cite{JegouDS11}, IVF\_PQ~\cite{JegouDS11}, and NSG~\cite{RNSGAli19}) are memory-centric and traverse their structures through raw pointers. Implementing them within the page storage framework forces every access through the page buffer, introducing page indirection, buffering overhead, and CPU cost a native library avoids. This overhead affects both search and update operations alike. 
A recent case study~\cite{VecDBRDBMSICDE24} measured this gap: a PostgreSQL-based implementation is several times slower than a native library under identical index structures and parameters. This page-versus-pointer mismatch persists even for disk-oriented indexes such as DiskANN~\cite{DiskANN19}: within PostgreSQL's storage engine, their in-memory components are still reached through the page buffer rather than directly, and can be evicted by heap pages that share the buffer pool.
Index updates face a further difficulty: vector indexes such as HNSW are inefficient to update in place (i.e., modifying the index structure directly), since in-place updates degrade index quality and are slow~\cite{FreshDiskANN, SPFresh23}. They are better maintained out-of-place, as in an LSM-style design~\cite{LSMTree,LuoC20,Milvus21}. However, PostgreSQL's storage engine makes out-of-place updates hard to adopt.

\myline{Engineering overhead}
Beyond runtime cost, re-engineering a memory-centric index into a fixed-size page layout demands substantial engineering effort~\cite{PASE20}. Pointer-based structures must be re-expressed as page and offset references, large vectors may span multiple pages, and the index logic must interact correctly with the buffer manager to ensure crash safety. This complexity increases implementation and maintenance burden and makes it harder to track and adopt the rapidly evolving landscape of vector indexes.
 
\myline{Difficulty exploiting emerging hardware}
Finally, binding the index to PostgreSQL's page structure and buffer manager makes it difficult to exploit emerging hardware. 
GPU-accelerated search~\cite{CAGRA2024,NVIDIAcuVS}, for example, requires the index to reside in GPU memory using a layout optimized for GPU kernels. In contrast, a page-oriented index requires deep integration with PostgreSQL, whose query-processing pipeline currently lacks mature GPU support.
 
\subsection{Decoupling the Vector Index}
\label{sec:bg-decouple}

In our earlier work~\cite{PostgreSQLV-CIDR26}, we introduced \cidr{}, which addresses the limitations above via a novel decoupling principle. Rather than fitting the index into a page-oriented layout, it implements the index as an independent component outside PostgreSQL's storage engine. It still conforms to \texttt{IndexAmRoutine}, so the query engine invokes it transparently. Decoupling lets the index be backed by a native library (e.g., Faiss or hnswlib) with direct memory access, eliminating overhead from the page buffer. It also makes the index pluggable, so that new index implementations can be substituted as the field advances. Decoupling further enables the index to adopt an LSM-style framework, the out-of-place update scheme favored by many specialized vector databases~\cite{Milvus21, SingleStoreV24}: new vectors are appended to an in-memory mutable memtable that is later sealed and built asynchronously into an immutable index segment, with smaller index segments periodically merged. Because updates modify only in-memory memtables and never touch existing segments, this scheme avoids the in-place updates that vector indexes handle poorly. For a top-$k$ query, the similarity search runs on every memtable and segment, and their top-$k$ results are merged into the final answer.

Decoupling, however, leaves the index without crash consistency. For a normal page-based index, PostgreSQL logs its modifications together with the heap's as WAL records and replays them after a crash, recovering the index to a state consistent with the heap.
In contrast, as the decoupled index lives outside the data page buffer, its updates generate no WAL records, so a crash can leave it out of sync with the heap: vectors in the memtable will be lost even if the heap committed them, while flushed segments may retain vectors the heap never committed.

\cidr{} resolves this by separating index {metadata} from index contents, tracking the metadata in a newly introduced structure called {status pages}. The metadata is small: for each active index entry, it records the entry's identifier together with the TID of the corresponding heap tuple. \cidr{} implements status pages as ordinary PostgreSQL pages, so they reside in the data page buffer and are logged and recovered by PostgreSQL's WAL mechanism, which guarantees that they stay consistent with the heap across crashes. The actual vector index structures remain in the decoupled component and are not logged. 
After a crash, PostgreSQL conducts its recovery procedure, which restores the status pages to a consistent state. They then serve as a single source of truth: by comparing the recovered metadata against the decoupled index, the system repairs the index, fetching missing vectors from the heap via their recorded TIDs and appending them to memtables, and invalidating entries that should not exist. In effect, the status pages reattach the decoupled index to PostgreSQL's crash-recovery guarantees without forcing the index itself into the page-oriented format.

\myline{Limitations of \cidr{}}
The decoupling principle is elegant, and \cidr{} demonstrates the feasibility of the design with promising performance results. However, it remains a prototype with three important gaps. First, the decoupled index is not designed for PostgreSQL's multi-process execution model: because the index cannot be shared across processes, \cidr{} lacks concurrency support and is confined to a single connection.
Second, because the status pages record every active entry in a flat layout, crash recovery must process all of them arbitrarily, so recovery time and the resulting unavailability window grow with the index size.
Third, the design provides no support for replication, on which production deployments rely for read scaling. 

\section{System Overview}
\label{sec:overview}

\begin{figure*}[t]
  \centering
  \includegraphics[width=1.0\textwidth]{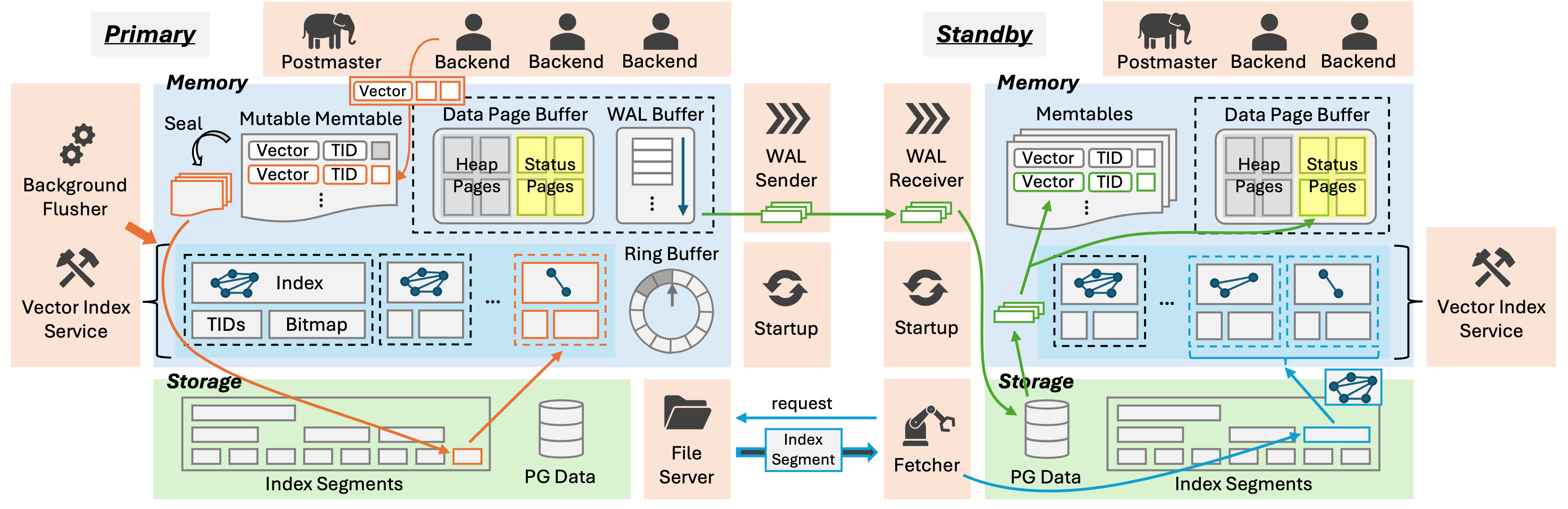}
  \caption{System Overview of \sys{}}
  \label{fig:sys}
\end{figure*}

Figure~\ref{fig:sys} shows the architecture of \sys{}, a vector database built inside PostgreSQL that can run in both single-node and multi-node settings. The figure illustrates the multi-node setting with a primary and a standby node. \sys{} is fully compatible with existing PostgreSQL-based vector search systems such as pgvector~\cite{Pgvector},  pgvectorscale~\cite{pgvectorscale}, and \cidr{}~\cite{PostgreSQLV-CIDR26}, sharing the same embedding storage, vector index registration via \texttt{IndexAmRoutine}, and hybrid search interface (Section~\ref{sec:bg-impl}). As an integrated vector database, \sys{} builds heavily on PostgreSQL's existing architecture, while extending it with new components to overcome the performance degradation identified in Section~\ref{sec:bg-limit} and address the limitations in \cidr{} (Section~\ref{sec:bg-decouple}).

\myline{Normal Processing}
\sys{} fits into PostgreSQL's architecture, where each client connection is served by a backend process under the \emph{postmaster}. While heap tables remain in PostgreSQL's data page buffer, \sys{} follows \cidr{} in adopting the decoupling approach and LSM-style index framework, managing vector indexes as memtables and immutable segments allocated outside the page buffer. As shown by the orange arrows in Figure~\ref{fig:sys}, a backend inserts a vector into the mutable memtable; once the memtable fills, it is sealed into an immutable memtable; a background flusher builds an index over the immutable memtable and persists it as an on-disk index segment; the \texttt{VISP} (\emph{Vector Index Service Process}) loads the resulting segment into its local memory. \sys{} removes \cidr{}'s single-connection limitation by introducing \texttt{VISP}, a dedicated process that exclusively owns all index segments, each consisting of an index, a TID array, and a bitmap. \texttt{VISP} serves search requests from all backends over a shared ring buffer, making the decoupled index fully concurrent for the first time (Section~\ref{sec:concurrency}).

\myline{Crash Recovery}
\sys{} leverages PostgreSQL's existing WAL and recovery infrastructure, shown in Figure~\ref{fig:sys} (the WAL buffer and the startup process that conducts log replay), by storing status pages in the data page buffer so they are logged and recovered alongside heap pages. \sys{} then layers a customized recovery procedure on top to restore the decoupled index, as \cidr{} also does. However, \sys{} redesigns the status pages and the recovery mechanism so that recovery time stays minimal and constant, in contrast to \cidr{}, whose recovery cost grows with the index size, becoming substantial at scale 
(Section~\ref{sec:crash_recovery}).

\myline{Physical Replication}
\sys{} supports physical replication in two paths, shown in Figure~\ref{fig:sys}: it extends PostgreSQL's replication mechanism with customized WAL log records and callback functions invoked during replay (green arrows) to propagate consistency-critical index operations to the standby, and introduces a file server and fetcher workers (blue arrows) to propagate index segments asynchronously. This gives \sys{} physical replication support that \cidr{} lacks entirely (Section~\ref{sec:replication}).

\section{Concurrency Support}
\label{sec:concurrency}

The first gap in \cidr{}~\cite{PostgreSQLV-CIDR26}  is the decoupled index's
inability to be shared across PostgreSQL's processes (identified in Section~\ref{sec:bg-decouple}), which confines it to a single
connection. \sys{} removes this limitation and makes the decoupled index fully
compatible with PostgreSQL's concurrency guarantees. Two properties are required:
multiple connections must be able to access the same index simultaneously
(Section~\ref{sec:index_worker} and ~\ref{sec:visp_concerns}), and concurrent transactions must remain correct
under PostgreSQL's isolation levels (Section~\ref{sec:concurrency_correctness}).

\subsection{Converting Multi-Process Concurrency  to Multi-Thread} \label{sec:index_worker}

Supporting multiple connections in \sys{} is non-trivial due to PostgreSQL's \emph{multi-process} execution model, where each connection runs in its own process (see Section~\ref{sec:overview}).

This execution model poses fundamental challenges to integrating vector indexes in a decoupled, plug-in fashion. All existing vector index libraries (e.g., Faiss~\cite{GithubFaiss}, HNSWlib~\cite{hnswlib}, Knowhere~\cite{knowhere}) assume a \emph{multi-threaded} concurrency model and rely on raw pointers within a virtual address space, which prevents direct sharing across PostgreSQL processes.

A straightforward approach is to implement a vector index that adopts a multi-process execution model. However, doing so not only introduces substantial engineering complexity but also undermines the flexibility benefits of decoupling, since it disables the modular reuse of existing vector indexing libraries.

Thus, we propose a new design that addresses this challenge at its root. We introduce the \emph{Vector Index Service Process} (\texttt{VISP}), a dedicated process that exclusively owns all vector index instances in its local memory and provides index service to all PostgreSQL processes.
The \texttt{VISP} manages every index segment, each consisting of a vector index, an array of TIDs that maps each indexed vector back to its heap tuple, and a deletion bitmap. Deletions within a segment are applied softly by flipping the corresponding bit in the bitmap rather than modifying its index structure directly. With these structures, the \texttt{VISP} executes vector searches on behalf of all backend processes. 

A vector search request flows from the issuing backend to the \texttt{VISP} and back. The backend enqueues its request into a shared ring buffer, the communication channel to the \texttt{VISP}. \texttt{VISP} drains the buffer and serves requests with a thread pool. This arrangement gives two levels of parallelism. Across requests, the ring buffer collects searches from all backends and the thread pool serves them concurrently, letting a single \texttt{VISP} handle many connections at once. Within a request, the search fans out across LSM segments in parallel, and their per-segment top-$k$ results are merged into the top-$k$ across segments before the issuing backend receives them.

A search must also cover the memtables, which hold the most recently inserted vectors (Section~\ref{sec:overview}). Because memtables store raw vectors contiguously as arrays rather than relying on pointer-based index structures, they reside in PostgreSQL's static shared memory and are directly accessible by all backend processes. Thus, while the \texttt{VISP} searches the flushed segments, the requesting backend concurrently scans the memtables for their top-$k$ candidates, then merges the memtable and segment results into the final output. The cost of scanning the memtables is therefore absorbed into the segment search rather than adding extra latency.

The \texttt{VISP} also runs maintenance tasks over flushed segments: loading newly flushed ones, merging smaller segments into larger ones~\cite{jin2026efficient}, and rebuilding a segment when the deletion ratio in the bitmap exceeds the threshold. These run in the background, off the critical path of foreground searches.

\subsection{More on The Vector Index Service Process}
\label{sec:visp_concerns}
Centralizing all vector-index operations in a single \texttt{VISP} may raise two natural concerns: that it is a single point of failure, and that it is a performance bottleneck. We address both.

\subsubsection{Single Point of Failure}
One might worry that a \texttt{VISP} crash leaves the index in an inconsistent state or renders vector search unavailable to all backends. We argue that the impact of such a crash is minimal, for two reasons. First, a \texttt{VISP} crash never corrupts existing index segments. Flushed segments are immutable once written: \texttt{VISP}'s foreground work only searches them, and its background merges and rebuilds produce new segments rather than modifying existing ones in place. All insertions, meanwhile, are confined to memtables managed by backend processes, not \texttt{VISP}.
Second, although a \texttt{VISP} crash also causes a termination and restart of all PostgreSQL processes and a subsequent crash recovery, this is general PostgreSQL behavior, triggered by the abnormal exit of any process attached to shared memory, not something specific to \texttt{VISP}: an ordinary process (e.g., a backend) crashing would cause the identical restart. 
However, \sys{} minimizes the resulting downtime by addressing its two costs separately: a lightweight crash-recovery mechanism (Section~\ref{sec:crash_recovery}) restores the index to a consistent state with negligible overhead, while the remaining cost, making the index \emph{searchable} again, is the cold-start problem we address next.

\myline{Cold-start availability}
\sys{}'s LSM-based framework answers every query by searching all index segments, which must reside in memory to be searched for indexes such as HNSW. After a restart, no segments are in memory. In the default mode, all segments are loaded in bulk, triggered by the first query against the index. The index is not searchable until the entire load finishes.
This is a sharper problem than slow queries: it is a window of search unavailability, not just increased latency. It also distinguishes \sys{} from page-oriented vector indexes within PostgreSQL's storage engine (e.g., pgvector), which fault in only the index pages they visit, amortizing the cost across queries so that search stays available, albeit slower at first. We emphasize that this window is not specific to \sys{}, but common to any vector database that adopts native vector index libraries, since they require in-memory state before they can be searched. A memory-centric index such as HNSW~\cite{HNSW18} resides entirely in memory, and even a disk-based index such as DiskANN~\cite{DiskANN19} has in-memory components that must be loaded first. The window can be substantial (we measured ~5s for an HNSW index of 10M vectors and ~43s for 100M, see Section~\ref{sec:ev_cold_start}).

We close this window with an \texttt{mmap}-based cold-start mode, exploiting the \texttt{mmap} support in many vector index libraries (e.g., Faiss~\cite{GithubFaiss}, Knowhere~\cite{knowhere}). \texttt{mmap} is an OS-provided mechanism that maps file contents directly into a process's virtual address space, allowing the file to be accessed through ordinary pointers. Pages load lazily on first access: touching a page not yet resident in memory triggers a page fault, causing the OS to read it from disk into the system page cache.
In the \texttt{mmap}-based cold-start mode, the index segments are mapped into the address space via \texttt{mmap} on the first query and become searchable right away, while \texttt{VISP} promotes segments to full in-memory residency in the background, loading entire segments into \texttt{VISP}'s local memory. This shifts the bulk of the loading cost off the critical path: early queries pay only the cost of page faults as the OS loads segment data on demand, so the index becomes searchable in nearly two orders of magnitude less time than loading entire segments upfront. Once the background promotion completes, subsequent queries are served from the pure in-memory index for better performance, and the memory mapping is released.

We are aware of the concerns raised by Crotty et al.~\cite{MMAP_CIDR}, who caution against using \texttt{mmap} for DBMS file I/O due to correctness and performance issues. The correctness concern, namely transactional safety risk, arises because the OS can flush dirty pages before a transaction commits, which stems from the DBMS having no control over how and when mapped pages are written back to disk. However, \sys{} maps only immutable, read-only segment files that are fully persisted before mapping, so no writes go through the mapping and no dirty pages or write-ordering races can arise. The correctness issue therefore does not apply to \sys{}'s use of \texttt{mmap}.
Beyond correctness, Crotty et al. also identify two performance hazards: I/O stalls from blocking page faults on pages not yet resident in the page cache, and \emph{TLB shootdowns}, where releasing a memory mapping forces the OS to interrupt every core to invalidate cached address translations. In \sys{}'s cold-start use, the I/O stalls from page faults are intentional: they replace the far worse outcome of complete search unavailability after a restart. The remaining hazard, TLB shootdowns from releasing mappings, is bounded because \sys{} releases the cold-start mapping only once, when background promotion to full in-memory residency completes. Importantly, the authors' guidance~\cite{MMAP_CIDR} permits mmap for read-only data used transiently rather than as a permanent buffer pool replacement, which matches \sys{}'s design precisely. \sys{} therefore scopes its use of \texttt{mmap} deliberately to avoid the risks above.

\subsubsection{Performance Bottleneck}
A second concern is that a single \texttt{VISP} could become a performance bottleneck. Although there is only one \texttt{VISP} process, it employs multiple threads that collectively leverage the server's CPU resources. Our experiments in Section~\ref{sec:ev_memorystaticsearch} show that throughput scales with increasing concurrency rather than saturating, indicating that the \texttt{VISP} is not the limiting factor. The actual bottleneck lies in the server's CPU and memory resources, which we address by adding support for physical replication (Section~\ref{sec:replication}).

\subsection{Concurrency Control in \sys{}} \label{sec:concurrency_correctness}

The second goal of \sys{} is to ensure the correctness of concurrent transactions under different isolation levels in PostgreSQL when the decoupled vector index is involved. This requires two properties: \emph{visibility correctness}, guaranteed entirely at the heap level and unaffected by the index, and \emph{physical correctness}, guaranteed by the index's own concurrency control over its structures.

\myline{Visibility correctness}
PostgreSQL employs MVCC, allowing multiple physical versions of a row (heap tuples) to coexist, with visibility determined by transaction-specific snapshots. Index scans, whether on built-in or decoupled ones, may return TIDs referencing dead or invisible versions. \sys{} preserves correctness by enforcing visibility checks at the heap level after fetching the corresponding tuples. Consequently, decoupling the vector index does not affect visibility correctness, which remains guaranteed by the heap.

\myline{Physical correctness}
Beyond visibility, the decoupled vector index ensures physical correctness under concurrent access. In \sys{}, concurrent reads and writes occur only on the mutable memtable given the properties of the LSM. To support scalable concurrency, we adopt a lock-minimal, append-only design. Writers reserve insertion slots in the memtable using lightweight synchronization primitives and write vector data without holding global locks. Once a write completes, a ready bit is atomically set. Concurrent scans check this bit and skip unready entries, preventing access to partially written data. This design guarantees correctness while keeping concurrent access efficient.

\section{Crash Recovery} \label{sec:crash_recovery}

The second gap in \cidr{}~\cite{PostgreSQLV-CIDR26} is recovery scalability: its crash recovery mechanism (as described in Section~\ref{sec:bg-decouple}) ties the recovery overhead to the total size of the index, so the larger the data grows, the longer the system stays unavailable after a crash. \sys{} addresses this gap by redesigning the status pages to exploit properties of the LSM framework and revising the recovery mechanism accordingly, keeping recovery cost bounded and independent of index size. We first identify why the original layout scales poorly (Section~\ref{sec:why_inefficient}), then present the redesigned status pages (Section~\ref{sec:status_redesign}), how they are maintained during normal processing (Section~\ref{sec:status_normal}), and how the index is recovered (Section~\ref{sec:index_recovery}).

\subsection{Why the Original Layout Fails to Scale}
\label{sec:why_inefficient}

The recovery cost of the original status pages in \cidr{}~\cite{PostgreSQLV-CIDR26} stems from the following two design choices.

\myline{Flat, segment-oblivious layout}
\cidr{}'s design does not exploit the properties of the LSM framework of the decoupled vector index. The original status pages record all index entries in a flat layout. However, tracking entries of flushed index segments serves no purpose, as they are immutable and durable on disk. Besides, the flat layout does not separate segment entries from volatile memtable entries, so recovery is forced to process the entire index rather than only the volatile portion that can actually be lost upon a crash.

\myline{Recording vacuum on durable segments}
The original status pages in \cidr{} also record modifications introduced by the vacuum procedure, which cleans up dead tuples left behind by MVCC, by physically removing their metadata from status pages. Since vacuum sweeps the whole index, including flushed segments, recovery must in turn process the whole index for those removals. We observe that this work is avoidable. Status pages exist to provide transactional durability: changes must be recoverable upon transaction commit. Vacuum, however, runs asynchronously outside any transaction, so its durability is independent of transaction boundaries. It suffices to flush the index vacuum changes to durable storage before returning to the index vacuum caller, which guarantees the correct durability ordering without routing vacuum through the status pages at all.

\subsection{Redesigning the Status Pages}
\label{sec:status_redesign}

\begin{figure}[tbp]
    \centering
    \includegraphics[width=0.9\linewidth]{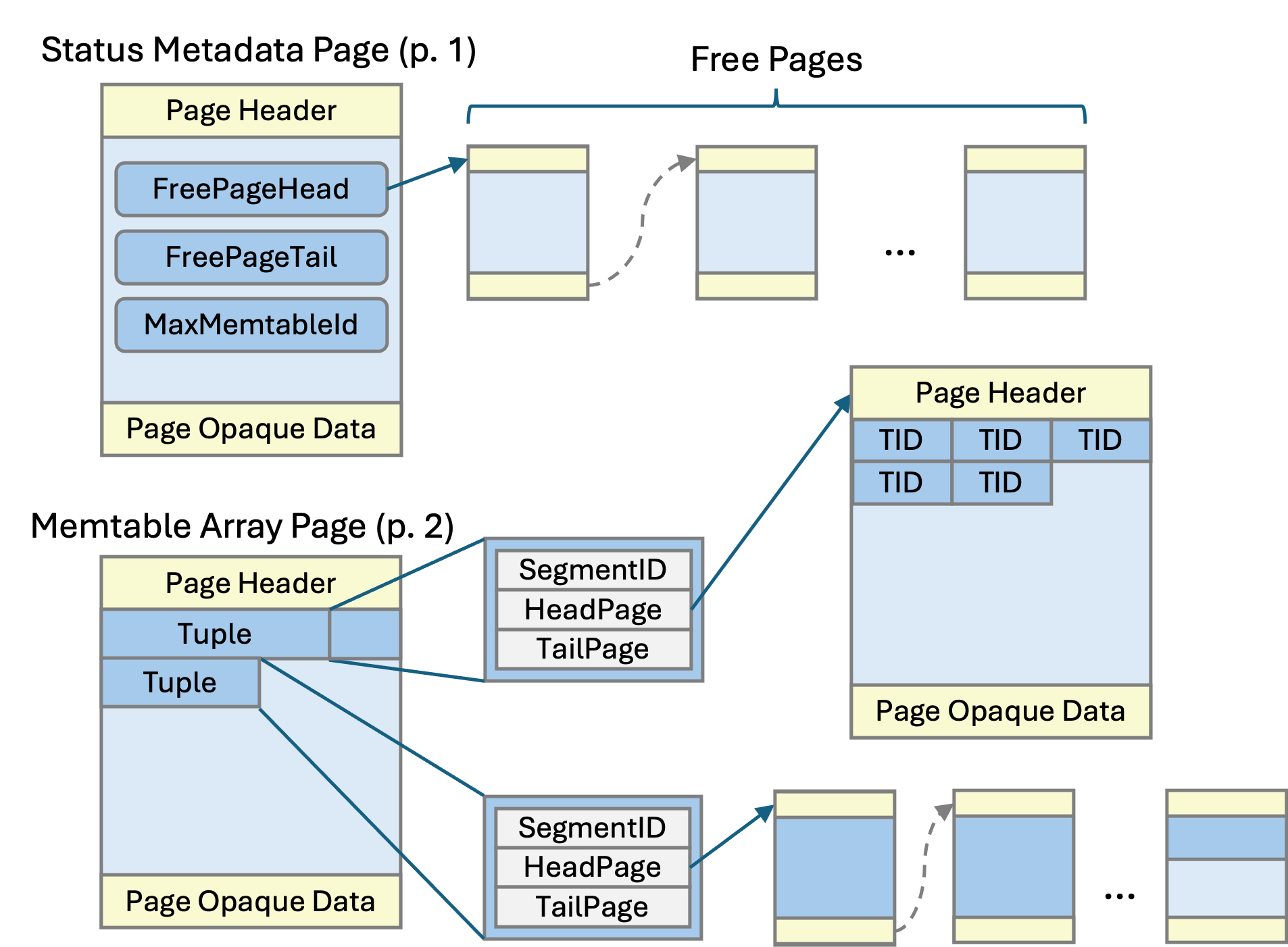}
    \caption{Status Page Design}
    \label{fig:status}
\end{figure}

To exploit the LSM structure, \sys{} replaces the original flat status-page layout with one organized by \texttt{SegmentID}. \texttt{SegmentID} is the identifier the LSM framework assigns to each memtable upon its creation, in monotonically increasing order. It continues to track the memtable's vectors throughout their lifetime, including across flushes, merges, and rebuilds.

Figure~\ref{fig:status} shows the resulting structure. Status pages record the TID of every active vector in the index, and organize these TID records by \texttt{SegmentID}. Since all vectors in a memtable share one \texttt{SegmentID}, their TIDs are stored contiguously across a page list of status pages, with the pointer to the next page kept in each page's opaque-data field (the fixed-size region at the end of a PostgreSQL page reserved for customized metadata). 
The \texttt{SegmentID} and its page list are recorded together in a dedicated \emph{memtable array page}: each tuple stores its \texttt{SegmentID} along with a \texttt{PageHead} and \texttt{PageTail} pointer locating the first and last page of that list, so new TIDs can be appended to the tail without traversing it.

Because flushed segments are immutable and durable, their entries no longer need tracking, so \sys{} removes the corresponding page list from the status pages once a memtable is flushed. Without reusing those pages, the status pages would grow without bound. 
Therefore, \sys{} adopts the free-page list design used by some built-in indexes in PostgreSQL, such as GiST~\cite{pggist}: the free-page list tracks reclaimed status pages so they can be reused by subsequent memtables, and its metadata is stored in the \emph{status metadata page}, as shown in the upper part of Figure~\ref{fig:status}. This page also holds \texttt{MaxMemtableId}, the \texttt{SegmentID} of the current mutable memtable.

\subsection{Normal Processing}
\label{sec:status_normal}

Status pages are updated together with the decoupled vector index whenever the index update operations defined in \texttt{IndexAmRoutine} are invoked by the PostgreSQL core engine. Index insertions add entries and vacuum removes them. We describe how insertions and vacuum maintain the new status pages.

\begin{sloppypar}
When a vector is inserted, the core engine writes the tuple to heap pages, logs it to the WAL, and invokes the index's insert operation, which appends the vector to the mutable memtable. \sys{} then appends the corresponding TID to the mutable memtable's page list in status pages. If the current memtable is full, the insert first seals it and creates a new mutable memtable to hold the vector. The status pages follow this by advancing \texttt{MaxMemtableId} in the status metadata page and registering the new memtable in the memtable array page. A new page list is assigned to the new \texttt{SegmentID}, taken from the free-page list when reclaimed pages are available or allocated fresh otherwise. Finally, when \sys{} detects during an insert that an immutable memtable has just been flushed, it removes that memtable from the status pages by clearing its tuple in the memtable array page and moving its page list to the free-page list, keeping the status pages bounded to the entries of unflushed memtables.
All of these status page updates generate WAL records, and both the heap's and the status pages' WAL records are flushed upon transaction commit, guaranteeing their durability and consistency after recovery.
\end{sloppypar}

Deletion in PostgreSQL is lazy: deleting a heap tuple only updates the MVCC metadata in its header to record the deletion, rather than removing the tuple. Query correctness is preserved by the heap-level visibility check applied after each search (Section~\ref{sec:concurrency_correctness}). 
Accordingly, \texttt{IndexAmRoutine} exposes no direct delete. 
Deleted tuples are removed once they are dead, that is, invisible to all active transactions. PostgreSQL's vacuum process periodically scans heap pages to collect all dead tuples and triggers index vacuum to clear their index entries before it removes the heap tuples. \sys{} clears each index entry depending on where it resides.

If it lies in a memtable, \sys{} removes its TID from the status pages, generating a WAL record, and marks the entry in the memtable's in-memory bitmap. If it lies in a flushed segment, \sys{} marks that segment's bitmap. At the end of the index vacuum, the affected segment bitmaps are force-flushed to disk, making them durable without involving status pages. 
A crash that occurs before the force flush completes may cause a bitmap update to be lost, leaving some index entries that vacuum was meant to remove still marked valid. However, this causes no inconsistency or correctness issue: heap vacuum only removes a tuple after its index vacuum completes, so no dangling entries can result, and these remaining entries, though pointing to dead tuples, are still filtered out by the heap-level visibility check (Section~\ref{sec:concurrency_correctness}), keeping search correct.

\subsection{Index Recovery}
\label{sec:index_recovery}

After a restart, PostgreSQL's \emph{startup process} first recovers the status pages by replaying WAL records, making them consistent with the heap. The decoupled vector index is then recovered based on the restored status pages. Because flushed memtables are removed from the status pages during normal processing, the recovered status pages describe only memtables that may not have been persisted before the crash, so, in the common case, the work of restoring the decoupled index reduces to reconstructing those memtables. Flushed segments are involved only in certain cases, illustrated by the example below.

Consider the following example. Suppose the memtable capacity is 10K vectors and the current mutable memtable (\texttt{SegmentID} 100) already holds 9K. A transaction then inserts 15K vectors, sealing and flushing memtables 100 and 101 and allocating a new memtable (\texttt{SegmentID} 102). If \sys{} crashes before this transaction commits, its WAL records may never reach disk, so after recovery the heap contains none of these vectors---neither the final 1K of memtable 100 nor any of memtables 101 and 102. The recovered status pages likewise keep only the first 9K vectors in memtable 100, omit memtables 101 and 102, and still record \texttt{MaxMemtableId} at 100. Yet segments 100 and 101 were already flushed to disk by the background flusher, making them durable. These flushed segments are therefore inconsistent with the heap and must be reconciled during recovery.

\sys{} begins by reading \texttt{MaxMemtableId} and dropping any flushed segment whose \texttt{SegmentID} exceeds it (segment 101 in the example above). It then iterates over the memtable array page, and for each \texttt{SegmentID}, scans its page list to recover the corresponding segment or memtable. If the \texttt{SegmentID} corresponds to a flushed segment (segment 100), \sys{} reconstructs that segment's bitmap, marking a vector valid if its TID appears in the status pages and invalid otherwise. For a merged segment spanning several memtables, only the entries for the given \texttt{SegmentID} are touched. 
If the \texttt{SegmentID} corresponds to no flushed segment, the memtable's vectors were never persisted and are lost, so \sys{} rebuilds the memtable by scanning the status pages for all TIDs of that \texttt{SegmentID} and re-fetching vectors from the heap, reinserting each with its TID.

\myline{Correctness of the recovery process} 
\sys{} recovery relies on two distinct durability boundaries: one marked by the recovered status pages, the other by which segments the LSM framework had flushed at crash time. The status page boundary is authoritative, since the status pages are durable through PostgreSQL's WAL mechanism and thus consistent with the heap. Because the two boundaries can lead or lag one another, matching the index to the status pages requires one of two corrections: \sys{} reconstructs the corresponding memtable from the heap where status page entries exceed flushed segments, or trims or discards flushed segments where they exceed the status page boundary. \sys{} performs exactly this correction at recovery time, so the resulting index reflects precisely the vectors present in the heap after recovery, no more and no less. This work is bounded by the small number of entries recorded in the status pages at the time of the crash, so the cost stays low and constant regardless of how large the index has grown. 

\section{Physical Replication}
\label{sec:replication}

Vector search in production is a read-heavy, high-volume workload: services such as RAG, semantic search, and recommendation issue a large number of vector similarity queries, each far more compute-intensive than a traditional point lookup. \emph{Physical replication} serves exactly this setting: a primary node streams its WAL to read-only standbys that replay it to track the primary while remaining queryable, letting them absorb the query fan-out of costly vector search. 
For a built-in PostgreSQL index, this support is free: every index change passes through the data page buffer and is logged to the WAL, so replaying that log on the standby reapplies the same page-level changes and catches up the index. The decoupled index sacrifices this for the same reason it forfeits crash consistency: its contents live outside the data page buffer and generate no WAL.

Blindly replaying the WAL on a standby in \sys{} only reproduces the index's status pages and the corresponding heap pages. The unique challenge in \sys{} is that, the index structures are never logged, leaving the standby with no searchable index to answer queries. \sys{} addresses this gap with a replication mechanism designed specifically for the decoupled index, described next.
 
\subsection{Two-Channel Design}
\label{sec:repl-design}

Replication must give the standby the same guarantee the decoupled design already gives on a single node: the index stays consistent with the heap. To meet this requirement, \sys{} splits the operations of the decoupled index by their effect on consistency. Our observation is that insertion and vacuum operations change which entries the index holds relative to the heap, so they are critical to consistency. In contrast, flushing, merging, and rebuilding operations only reorganize vectors already in the index into a different physical representation (i.e., the number of segments and the index structure of each segment), so they govern how the index is searched rather than which entries it holds, thus affecting performance but not correctness. We call the first class \emph{consistency-critical operations} and the second \emph{performance-critical operations}.

The two classes therefore impose very different replication requirements. Consistency-critical operations carry a correctness requirement: they must be durably replicated to the standby, and applied there in the same order as the heap changes they accompany on the primary, or the index will disagree with the heap. Performance-critical operations carry only a best-effort requirement: missing or reordering the propagation of index segment files generated by flushing, merging, and rebuilding is acceptable. \sys{} meets these with two different replication paths (Figure~\ref{fig:sys}): a \emph{consistency-critical channel} (the green arrows) for consistency-critical operations and a \emph{performance-critical channel} (the blue arrows) for the segment files. 

\myline{Consistency-critical channel} \sys{} replicates insertion and vacuum operations based on PostgreSQL's extensibility framework. By default, WAL records capture only physical changes to PostgreSQL's pages, so replaying them recovers nothing beyond those pages. However, the extension framework allows attaching a custom record type and arbitrary payloads to WAL records, and registering a callback for each record type that runs on replay. This callback can operate on components other than PostgreSQL's pages, such as the decoupled index, which PostgreSQL does not otherwise manage. \sys{} uses this framework to define its own record types for the decoupled vector index, each carrying the payload and callback it needs. These WAL records are generated, shipped, and replayed alongside all others in a single ordered stream, preserving their order relative to the heap changes they accompany.

\sys{} defines three customized record types that together cover every consistency-critical change to the decoupled index: \texttt{insert}, \texttt{memtable\_creation}, and \texttt{vacuum}. Since each of these changes modifies status pages through the data page buffer, PostgreSQL already generates WAL records for them. \sys{} extends these records with a custom type, payload, and replay callback tailored to each case.

The WAL record generated when an insertion writes a TID to status pages (Section~\ref{sec:status_normal}) carries an \texttt{insert} record type and both the TID and vector embedding as payload. Carrying the embedding with the TID lets the standby append the vector into its memtable directly from the WAL, without a heap lookup. Similarly, the WAL record generated when a new memtable is registered carries a \texttt{memtable\_creation} type and the new memtable's \texttt{SegmentID}, so the standby can open a matching memtable of its own.

Vacuum on a memtable entry removes a TID from a status page, and the resulting WAL record carries a \texttt{vacuum} record type with that TID as payload. Vacuum on a flushed-segment entry, however, follows a different path: it only flips a bit in the segment's bitmap, leaving status pages unchanged and generating no WAL record to carry a type. Single-node recovery tolerates this because it relies on the forced bitmap flush rather than status pages (Section~\ref{sec:status_normal}), but a remote standby never observes that flush. \sys{} therefore logs a standalone \texttt{vacuum} record carrying the TID, independent of any status-page write, so the deletion still reaches the standby.

\myline{Performance-critical channel}
A potential approach is that, index segment files could likewise be embedded in the WAL as payload, riding the same replication transport as the records above, but at a cost that outweighs the convenience. Flushing, merging, and rebuilding are background operations that the LSM framework keeps off the critical path, whereas the WAL is a single ordered stream that every commit must flush through and every replica must replay in order. Injecting large segment writes into this stream risks stalling commits on the primary, since a commit's durability requires flushing the WAL up to its own position, and any large record ahead of it must be flushed first. On the standby, replay faces the same ordering constraint: a large segment record blocks every record queued behind it, increasing replication lag. Besides, these records would also be replayed during PostgreSQL's recovery procedure after a crash, which inflates recovery overhead.

\sys{} therefore moves segment files onto a separate channel, while still emitting a small \texttt{segment\_create} WAL record for each segment through the WAL stream to notify the standby that a new segment exists, carrying only the segment's metadata as payload. This channel is served by fetcher workers on each standby, which extract the metadata from replayed \texttt{segment\_create} records and asynchronously send pull requests for the corresponding segment files to a lightweight file server on the primary. The file server, in turn, retrieves the requested segments from local storage and transfers them to the standby (Figure~\ref{fig:sys}). In this way, the heavy bytes stay off both the commit path and the replication stream.

Moving segments out of band opens architectural choices that routing them through the WAL would prevent. One such choice is running file servers independently of the primary node. A file server accesses only on-disk segment files and communicates only with fetcher workers, with no attachment to any other PostgreSQL component or process. This independence means that if segment files are also stored on other servers, file servers can be deployed there to serve segment requests from those locations, avoiding the network bottleneck that arises when multiple standbys pull large segment files from the primary simultaneously. Going further, \texttt{VISP} can be deployed on dedicated servers separate from PostgreSQL on both the primary and standby sides, so that index segment propagation proceeds directly between \texttt{VISP} nodes. None of this is possible if segment files instead travel through the WAL, which would require the PostgreSQL server to participate in every segment propagation, reintroducing the primary and standby PostgreSQL nodes into the transfer path and adding redundant network traffic.

The two channels therefore carry different weight: correctness rests entirely on the consistency-critical channel, which replays consistency-critical operations in strict order, while the performance-critical channel governs search efficiency by fetching new index segment files asynchronously.

\subsection{Replaying the Vector Index on the Standby}
\label{sec:repl-reconstruct}

To derive a searchable index from replication, \sys{} registers custom redo callbacks for both channels: handlers for the consistency-critical channel rebuild memtables and deletion bitmaps, while the \texttt{segment\_create} handler schedules an asynchronous fetch of the segment file over the performance-critical channel.

\myline{Reconstructing standby memtables}
An \texttt{insertion} record's callback appends the inline embedding payload directly to the standby memtable, while a \texttt{memtable\_creation} record's handler opens a new one. The standby never flushes, so these memtables accumulate, each serving as the correctness fallback for its vectors until the segment that covers it has been fetched and adopted.

\myline{Fetching and adopting segments}
Replaying a \texttt{segment\_create} record does not fetch the segment inline, instead, it schedules an asynchronous fetch in the fetcher process and lets replay continue. Once the segment's files arrive, the standby adopts it into the searchable index: any standby memtables it covers are freed, since their vectors are now durably represented in the segment, and any older segments it supersedes, from a merge or rebuild, are replaced. A late or out-of-order transfer may deliver an already-stale segment, identified by an outdated version number for a rebuild or a superseded \texttt{SegmentID} range for a merge, and discarded rather than adopted.

\myline{Tracking deletion}
Every \texttt{vacuum} record marks the deleted vector's bit in the relevant memtable's or segment's bitmap upon replay. Upon segment adoption, it is tempting to let the fetched segment carry its deletion bitmap and adopt that directly. But the side-channel is asynchronous, so the bitmap can lag the vacuums the logical channel has already applied and miss the latest deletions. Adopting it would silently resurrect deleted vectors. The standby therefore maintains its own deletion bitmaps from replayed vacuum records alone, never trusting the side-channel. When it adopts a new segment, it builds that segment's bitmap from the bitmaps already held in the memtables and segments the new segment supersedes.

\section{Experiments}
\label{sec:exp}
All experiments are conducted on Linux servers (Ubuntu 22.04, kernel 5.15) with two Intel Xeon sockets (112 cores total, 2 NUMA nodes) running at 2.0 GHz, 251 GB of DRAM, a 1.8 TB SSD, and a 10 Gb Ethernet NIC.

\subsection{Static Vector Search}
\label{sec:ev_staticsearch}
Static vector search measures latency and throughput for pure queries against a built index with no updates. We evaluate \sys{} first over memory-resident indexes (Section~\ref{sec:ev_memorystaticsearch}) and then over disk-resident indexes (Section~\ref{sec:ev_diskstaticsearch}). In both settings, we confirm that \sys{} preserves the latency advantage of a decoupled index over the page-oriented index, an advantage \cidr{}~\cite{PostgreSQLV-CIDR26} established. We then go further and show that this advantage extends to throughput under concurrency, a setting the \cidr{} prototype could not support at all.

\subsubsection{In-memory Vector Search} 
\label{sec:ev_memorystaticsearch}

\begin{figure*}[t]
  \centering
  \includegraphics[width=1.0\textwidth]{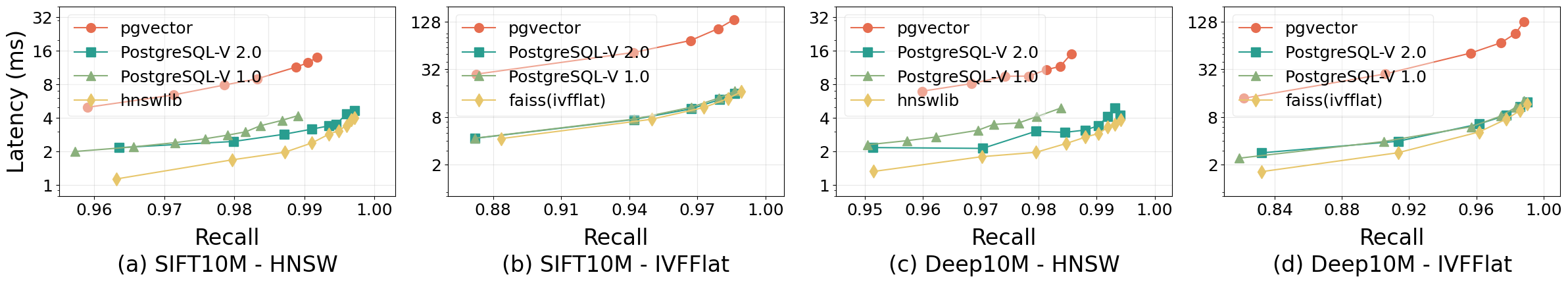}
  \caption{Recall-Latency Trade-off (In-memory)}
  \label{fig:ev_in_memory_latency}
\end{figure*}

\begin{figure*}[t]
  \centering
  \includegraphics[width=1.0\textwidth]{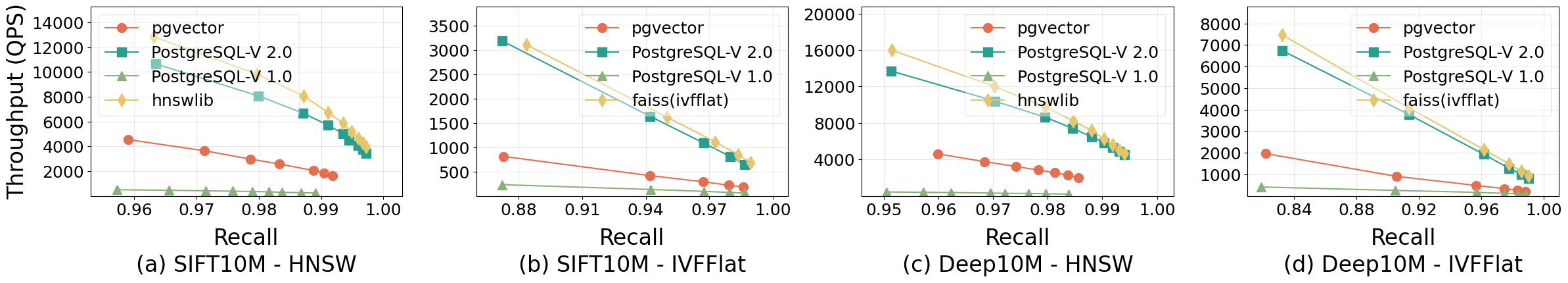}
  \caption{Recall-Throughput Trade-off Under Concurrency (In-memory)}
  \label{fig:ev_in_memory_thp}
\end{figure*}

We begin with the memory-resident indexes, comparing \sys{} against pgvector~\cite{Pgvector} configured with \hnsw{} and \ivfflat{}, as well as the \cidr{} prototype.  
To position \sys{} relative to specialized vector search systems, we additionally run it against the native libraries hnswlib~\cite{hnswlib} and Faiss~\cite{GithubFaiss}. \sys{} also adopts these libraries as its own decoupled index for \hnsw{} and \ivfflat{} respectively.
Every system is evaluated with matching index types and identical build-time parameters (\texttt{M=16}, \texttt{efConstruction=40} for \hnsw{}; \texttt{nlist=3162} for \ivfflat{}), so any performance difference reflects how the index is integrated rather than the index itself. For pgvector, we configure the buffer pool large enough to hold both the heap and its index in memory, and we warm up the cache before the evaluation. 
We report single-client search performance first and then turn to the concurrent, multi-client setting.

Under a single client, we evaluate the recall-latency trade-off on SIFT10M~\cite{SIFTData} and DEEP10M~\cite{YandexL16}, each holding 10 million vectors and 10,000 queries (Figure~\ref{fig:ev_in_memory_latency}). Backed by hnswlib, \sys{} is \textbf{3.2--4.4$\times$} faster than pgvector's \hnsw{} on SIFT10M and \textbf{3.2--5.0$\times$} faster on DEEP10M; backed by Faiss's \ivfflat{}, the gap widens to \textbf{6.5--8.6$\times$} over pgvector's \ivfflat{} on SIFT10M and \textbf{4.9--10.4$\times$} on DEEP10M.
Across both index types, \sys{} tracks the search performance of the standalone library it wraps, separated only by a small constant gap of roughly 0.8~ms spent parsing the SQL statement and fetching heap tuples for the TIDs the index returns, consistent with the gap reported by \cidr{}. 
Compared to the \cidr{}, \sys{} achieves comparable single-client latency on both datasets for \ivfflat{}. For \hnsw{}, \sys{} is slightly faster, as it adopts a newer version of the underlying \hnsw{} library. These results confirm that \sys{} preserves the decoupling benefit: the vector index is searched via direct memory access, avoiding the overhead that a page-oriented index incurs during search.

We then raise the client count to 32 to assess scalability, exercising concurrency that the \cidr{} could not support. Figure~\ref{fig:ev_in_memory_thp} reports recall versus aggregate throughput. 
Scaling from one client to 32, \sys{}'s HNSW throughput increases by \textbf{18.3$\times$ to 39.0$\times$} (recall spanning 0.96--0.99) and \ivfflat{} throughput by \textbf{10.7$\times$ to 13.7$\times$} (recall spanning 0.87--0.98). At 32 clients, throughput for both indexes matches the native libraries.
\ivfflat{}'s scaling is weaker than HNSW's because memory bandwidth becomes the bottleneck as concurrency increases, particularly at higher recall, where larger \texttt{nprobe} values raise memory traffic. This limitation comes from the index structure itself rather than \sys{}'s design. 
\sys{} also remains well ahead of pgvector: it delivers \textbf{3.2$\times$ and 3.6$\times$} higher throughput than pgvector's \hnsw{} and \textbf{3.4$\times$ and 4.2$\times$} higher than its \ivfflat{} across the two datasets.
Because \cidr{} restricts search to a single client, we compare against \cidr{} at its single-client maximum: \sys{} improves HNSW throughput by \textbf{23.4$\times$ to 36.4$\times$} and \ivfflat{} throughput by \textbf{11.1$\times$ to 16.2$\times$}, demonstrating the practical benefit of \sys{}'s concurrency support.

\subsubsection{Disk-based Vector Search}
\label{sec:ev_diskstaticsearch}

\begin{figure}[tbp]
    \centering
    \includegraphics[width=1\linewidth]{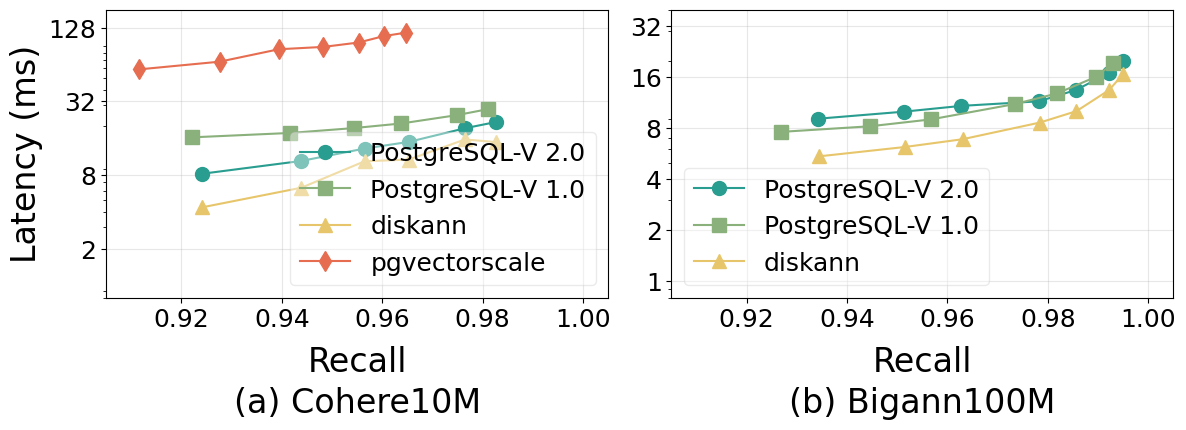}
    \caption{Recall-Latency Trade-off (Disk-based)}
    \label{fig:ev_disk_latency}
\end{figure}

\begin{figure}[tbp]
    \centering
    \includegraphics[width=1\linewidth]{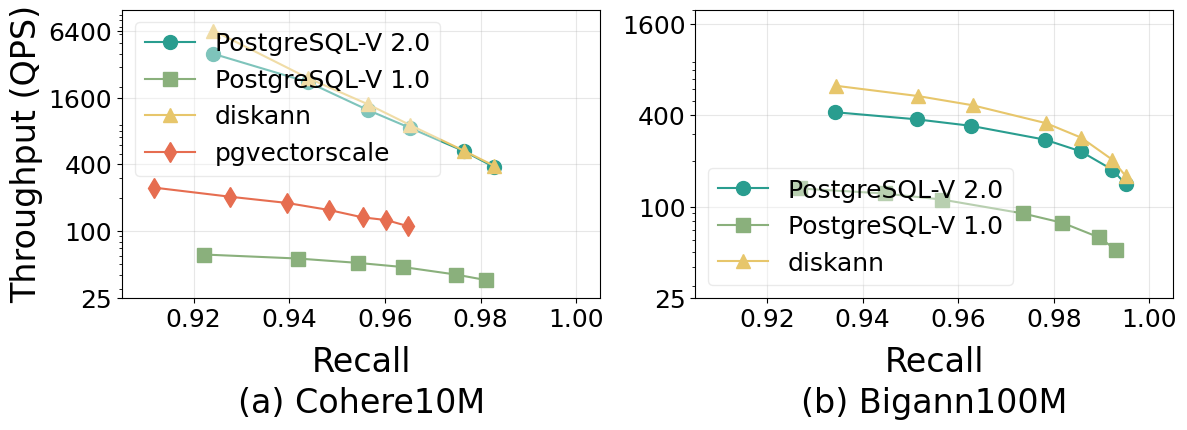}
    \caption{Recall-Throughput Trade-off Under Concurrency (Disk-based)}
    \label{fig:ev_disk_thp}
\end{figure}

\begin{sloppypar}
For the disk-resident indexes, \sys{} backs its decoupled index with the native DiskANN library~\cite{DiskANNGithub}. pgvector offers no disk-optimized index algorithm, so the baseline is pgvectorscale~\cite{pgvectorscale}, which ports a DiskANN-style index into PostgreSQL's page-oriented storage engine. We also compare against \cidr{}, likewise backed by DiskANN, and additionally run \sys{} against the native DiskANN library to bound the achievable performance.  To keep the comparison fair, every system is capped at 5~GB of memory. Each index uses its default parameters: \texttt{R=68}, \texttt{L=75}, \texttt{$\alpha$=1.2} for DiskANN, and \texttt{search\_list\_size=100}, \texttt{num\_neighbors=50}, \texttt{$\alpha$=1.2} for pgvectorscale. For DiskANN's beamwidth, each system uses the value that gives its best performance on each run. We report single-client search latency and, separately, peak throughput driven to the point where disk bandwidth saturates.
\end{sloppypar}

\begin{sloppypar}
On the Cohere10M dataset~\cite{Cohere} of 768-dimensional vectors, \sys{} (DiskANN) outperforms pgvectorscale by \textbf{7.3$\times$--8.0$\times$} in query latency (Figure~\ref{fig:ev_disk_latency}a) and \textbf{7.7$\times$--19.4$\times$} in peak throughput (Figure~\ref{fig:ev_disk_thp}a), primarily attributable to the direct control over which data resides in memory allowed by decoupling. \sys{} also tracks the native DiskANN library closely in both metrics. Against \cidr{}, \sys{} leads in latency by a narrow margin, a gap due to the different DiskANN library versions the two systems build on; in throughput, \cidr{} is measured at the single client it supports, as in the in-memory evaluation, and against that baseline \sys{}'s peak throughput is \textbf{10.5$\times$--65.3$\times$} higher.
\end{sloppypar}

We further scale to SIFT100M~\cite{BigANN}, of 100 million vectors. Here index construction in pgvectorscale exceeds a week and is no longer a practical baseline, so we compare \sys{} against the native DiskANN library and against \cidr{}. As Figures~\ref{fig:ev_disk_latency}b and~\ref{fig:ev_disk_thp}b show, \sys{} (DiskANN) stays close to the native library in both latency and throughput. Against \cidr{}, \sys{}'s single-client latency nearly matches and its peak throughput is \textbf{3.2$\times$--3.4$\times$} higher, a smaller gap than on Cohere10M, since DiskANN-backed systems saturate disk bandwidth at lower concurrency on SIFT100M, leaving \sys{}'s concurrency support less headroom to exploit.

\subsection{Dynamic Vector Search}
\label{sec:ev_updates}

\begin{figure}[tbp]
    \centering
    \includegraphics[width=0.95\linewidth]{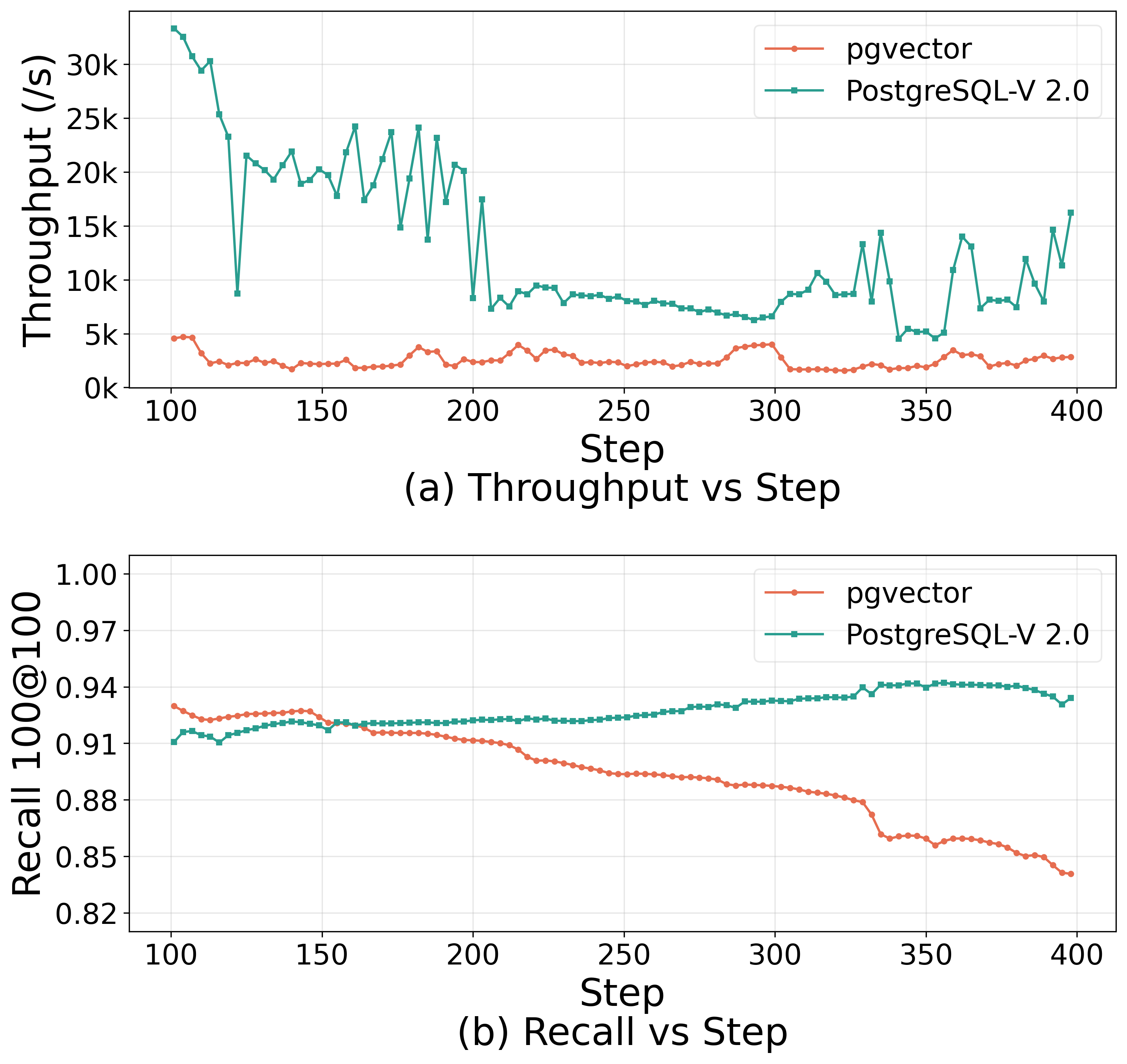}
    \caption{Dynamic Vector Search Results}
    \label{fig:ev_runbook_32_thp_recall}
\end{figure}

\sys{} fully implements the LSM-style framework of the decoupled index, including background segment merge and rebuild, which the \cidr{} prototype~\cite{PostgreSQLV-CIDR26} leaves incomplete. This lets us evaluate \sys{}'s dynamic vector search performance against pgvector, an evaluation \cidr{} could not support. 

We follow the evaluation methodology of the NeurIPS 2023 Big-ANN Competition~\cite{simhadri2024resultsbigannneurips23} to assess dynamic vector search performance using \emph{runbooks}~\cite{neurips2023streaming}. A runbook is an ordered sequence of batched insert, delete, and search operations executed step by step to emulate a streaming workload in a vector database.

\begin{sloppypar}
Specifically, we use the \texttt{msturing-10M\_slidingwindow} runbook~\cite{neurips2023streaming}, which consists of 400 steps. The first 100 steps insert half the data points (5M) in batches. The remaining 300 steps form 100 repeating three-step cycles: each cycle performs 10K vector searches, then deletes the oldest 50K vectors in the dataset, then inserts 50K new vectors, so the total dataset size stays constant throughout.
\end{sloppypar}

Note that these steps can be executed sequentially, with concurrency only within each step. However, in practice, queries, insertions, and deletions often occur simultaneously. Therefore, we execute the runbook in mixed mode, interleaving operations in three consecutive steps and issuing all operation types concurrently. In this setting, recall is difficult to compute because the ground truth is not well defined~\cite{zhang2025cleannefficientdynamismgraphbased}. Thus, we also conduct experiments in non-mixed mode and report recall results based on that setting. We perform experiments with 32 clients and measure throughput and the recall. Since the vector index is built after completing step 100, our evaluation begins at step 101.
Both \sys{} and pgvector use \hnsw{}, and as in the in-memory evaluation (Section~\ref{sec:ev_memorystaticsearch}), pgvector's buffer pool is sized to hold both the heap and its index.

\begin{sloppypar}
Figure~\ref{fig:ev_runbook_32_thp_recall} plots throughput and recall across the 300 steps. \sys{} sustains substantially higher throughput than pgvector throughout the run, averaging about \textbf{13.0K ops/s} against \textbf{2.5K ops/s} for pgvector, roughly a \textbf{5.1$\times$} gap overall, ranging from \textbf{3.4$\times$} to \textbf{8.5$\times$} across the run.
\end{sloppypar}

On recall, the two diverge over the run. \sys{}'s recall starts about 2\% below pgvector's but quickly catches up within 50 steps; pgvector then degrades steadily from \textbf{0.9298} to \textbf{0.8407}, whereas \sys{} climbs from \textbf{0.9106} to \textbf{0.9422}. pgvector's recall declines over successive delete-insert cycles, a known difficulty for in-place HNSW updates, where continuous deletions gradually degrade graph connectivity~\cite{FreshDiskANN,DagstuhlVectorSeminar26}. 
\sys{} avoids this because it never mutates a built index in place: deletions are recorded as soft deletes in per-segment bitmaps, and each segment's index is built once over an immutable vector set. 
Together, these results show that \sys{}'s out-of-place, LSM-style design sustains high throughput under continuous updates while avoiding the recall degradation that in-place index maintenance incurs.

\subsection{Crash Consistency and Recovery Overhead}
\label{sec:ev_crash_recovery}

We first test crash consistency, following the crash-injection methodology of~\cite{PostgreSQLV-CIDR26}. During a mixed-mode runbook execution, we kill the PostgreSQL process at a random point and clear the system page cache, then restart and compare the heap table against the recovered vector index to verify index consistency. Across more than 1,000 such trials, the recovered index agreed with the heap in every case, with no inconsistency detected. More detailed correctness analysis can be found in Section~\ref{sec:index_recovery}. 

We then measure recovery overhead, comparing \sys{} against pgvector and \cidr{}. The recovery overhead comprises the elapsed time of PostgreSQL's built-in recovery procedure -- during which the system scans and replays WAL records -- as well as the overhead introduced by our customized recovery mechanism for restoring the decoupled vector index. To substantiate the claim that the recovery time of the decoupled vector index is independent of the number of vectors stored, we configure the runbook sliding window to two different sizes: 10M and 100M. 
PostgreSQL's built-in recovery cost scales with the volume of WAL replayed, which is affected by PostgreSQL configurations and the operation rate. Therefore, across all three systems, we fix PostgreSQL configurations that affect WAL volume identically, including \texttt{max\_wal\_size} (1 GB), \texttt{min\_wal\_size} (80 MB), \texttt{wal\_level} (\texttt{replica}), \texttt{full\_page\_writes} (\texttt{on}), and \texttt{wal\_compression} (\texttt{off}). To control for operation rate, we drive \sys{} and pgvector with 16 threads each throttled to roughly 50 operations per second. \cidr{} supports neither concurrency nor full index update support, so it cannot be driven at this same controlled rate, making its PostgreSQL-level recovery cost incomparable to the other two systems; we therefore exclude it and compare only the overhead of its customized index recovery.

\begin{figure}[tbp]
    \centering
    \includegraphics[width=1\linewidth]{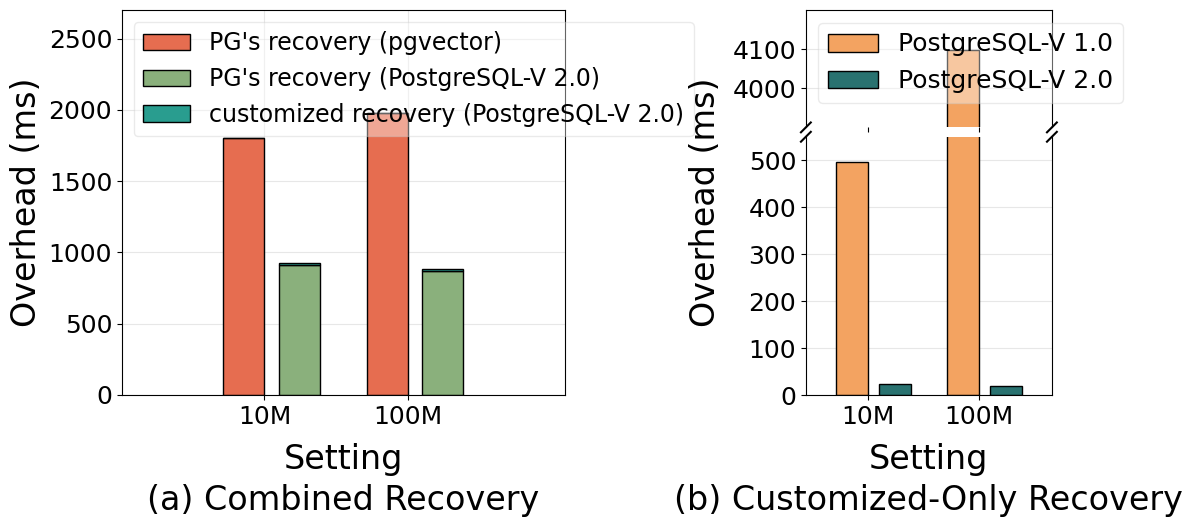}
    \caption{Recovery Overhead Evaluation Results} 
    \label{fig:ev_recovery}
\end{figure}

The results are presented in Figure~\ref{fig:ev_recovery}. \sys{}'s customized recovery mechanism averages \textbf{22.43 ms} and \textbf{19.76 ms} for sliding window sizes of 10M and 100M, respectively, negligible compared with its PostgreSQL built-in recovery overhead of approximately 900 ms. 
Moreover, as shown in Figure~\ref{fig:ev_recovery}a, PostgreSQL's recovery overhead in \sys{} is only about half of that in pgvector. This is because \sys{} generates less WAL volume during normal processing: \sys{} never logs modifications to its native index structures in the WAL. Its WAL consists solely of status-page WAL records, extended with insertion and vacuum payloads (Section~\ref{sec:status_normal}); in contrast, pgvector logs every modification to its vector index pages. Consequently, the WAL scanned during recovery is far smaller for \sys{} than for pgvector: approximately \textbf{200 MB} against \textbf{900 MB}.

Figure~\ref{fig:ev_recovery}b isolates \sys{}'s customized index-recovery cost against that of \cidr{}. \cidr{}'s index-recovery time tracks index size, rising from \textbf{495.3~ms} at 10M to \textbf{4,099~ms} at 100M, since that design scans every status page at recovery. The redesigned status pages remove this dependence: \sys{}'s index-recovery time stays near \textbf{20~ms} at both scales.

\subsection{Cold-Start Availability}
\label{sec:ev_cold_start}

This experiment evaluates the cold-start availability concern from Section~\ref{sec:visp_concerns}: after a restart, \sys{} must reload all segments of an index into the \texttt{VISP} before search can resume, creating a window of unavailability absent from page-oriented systems. 
We quantify this window against pgvector, and show that \sys{}'s \texttt{mmap} cold-start mode shrinks it dramatically relative to \sys{} without it. In our setup, both systems use \hnsw{}, and 16 clients issue vector search queries continuously against the SIFT10M dataset~\cite{SIFTData}, each at a rate of one query per 100~ms with a 100~ms timeout. 
We tune \texttt{ef\_search} for recall@100 of roughly 0.95.
During the run, the process under load is killed with \texttt{SIGKILL}: a backend process for pgvector, the \texttt{VISP} for \sys{}. In both cases, PostgreSQL's postmaster detects the abnormal exit of a process attached to shared memory, restarts the cluster, and triggers the crash recovery procedure. The recovery procedure is evaluated in Section~\ref{sec:ev_crash_recovery}, and this experiment focuses on what happens after recovery completes, before search can resume.

\begin{figure}[tbp]
    \centering
    \includegraphics[width=1\linewidth]{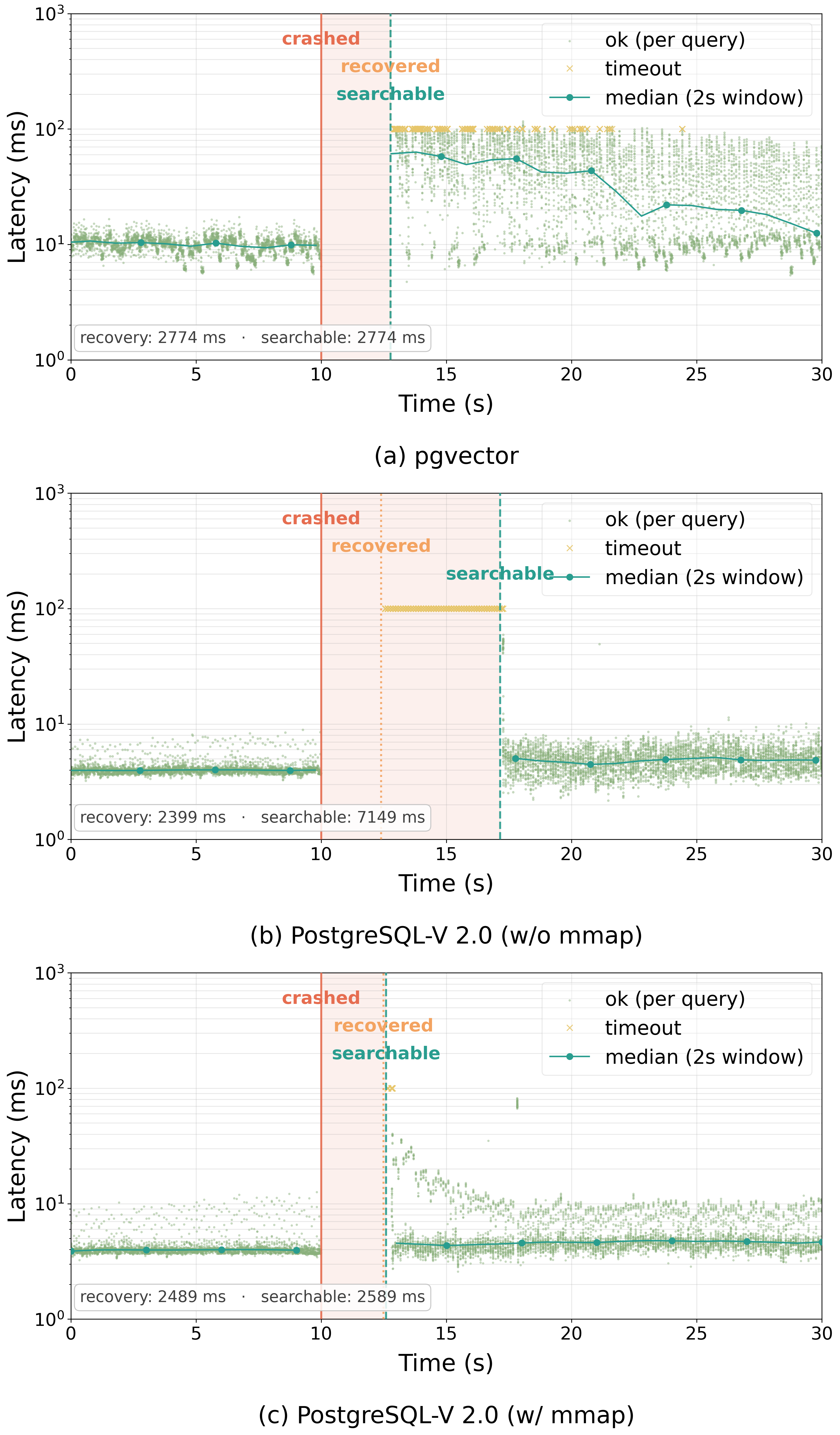}
    \caption{Query Latency During Server Restart}
    \label{fig:ev_restart_all}
\end{figure}

Figures~\ref{fig:ev_restart_all}a--\ref{fig:ev_restart_all}c show per-query latency over a 30-second window centered on the crash (crash at $Time = 10$~s). Each plot marks the crash, the moment connections are re-accepted (recovered), and the moment search resumes (searchable).

The result on pgvector (Figure~\ref{fig:ev_restart_all}a) is as expected: pgvector recovers in $\sim$2.8~s (crashed-to-recovered) and becomes searchable immediately. This is because it stores the index in PostgreSQL's pages and accesses them through the page buffer, so index pages are loaded into the buffer on demand and no upfront load is required. The total crashed-to-searchable time therefore remains $\sim$2.8~s, though the initial queries pay the cost of cold page-cache misses and their latency is visibly elevated until the working set warms up.

\sys{} without the \texttt{mmap}-based cold-start mode (Figure~\ref{fig:ev_restart_all}b) recovers in $\sim$2.4~s (crashed-to-recovered), but its decoupled index bypasses the page buffer and must be fully loaded into \texttt{VISP}'s private memory before search resumes: all segments load on the first search after recovery, so search stays unavailable until that load completes. This adds a substantial unavailability window of $\sim$4.8~s, bringing the total crashed-to-searchable time to $\sim$7.2~s.

\sys{} with the \texttt{mmap}-based cold-start mode (Figure~\ref{fig:ev_restart_all}c) recovers in $\sim$2.5~s (crashed-to-recovered), and instead of loading segments eagerly, it maps them into the \texttt{VISP}'s address space on the first query, making the index searchable roughly 100~ms after recovery, compared to $\sim$4.8~s without \texttt{mmap} (about 2\% of the no-mmap loading time). This brings the total crashed-to-searchable time to roughly 2.6~s. Background promotion to full in-memory residency completes in roughly 4.9~s, after which queries are served from the pure in-memory index and all \texttt{mmap} mappings are released.

To check whether the \texttt{mmap} cold-start mode's benefit holds at larger scale, we repeat the experiment on SIFT100M. At this scale, the recovered-to-searchable time is $\sim$43~s without \texttt{mmap}, while \texttt{mmap} reaches searchable in $\sim$670~ms.

\subsection{Physical Replication}
\label{sec:ev_replication}

\begin{figure}[tbp]
    \centering
    \includegraphics[width=1\linewidth]{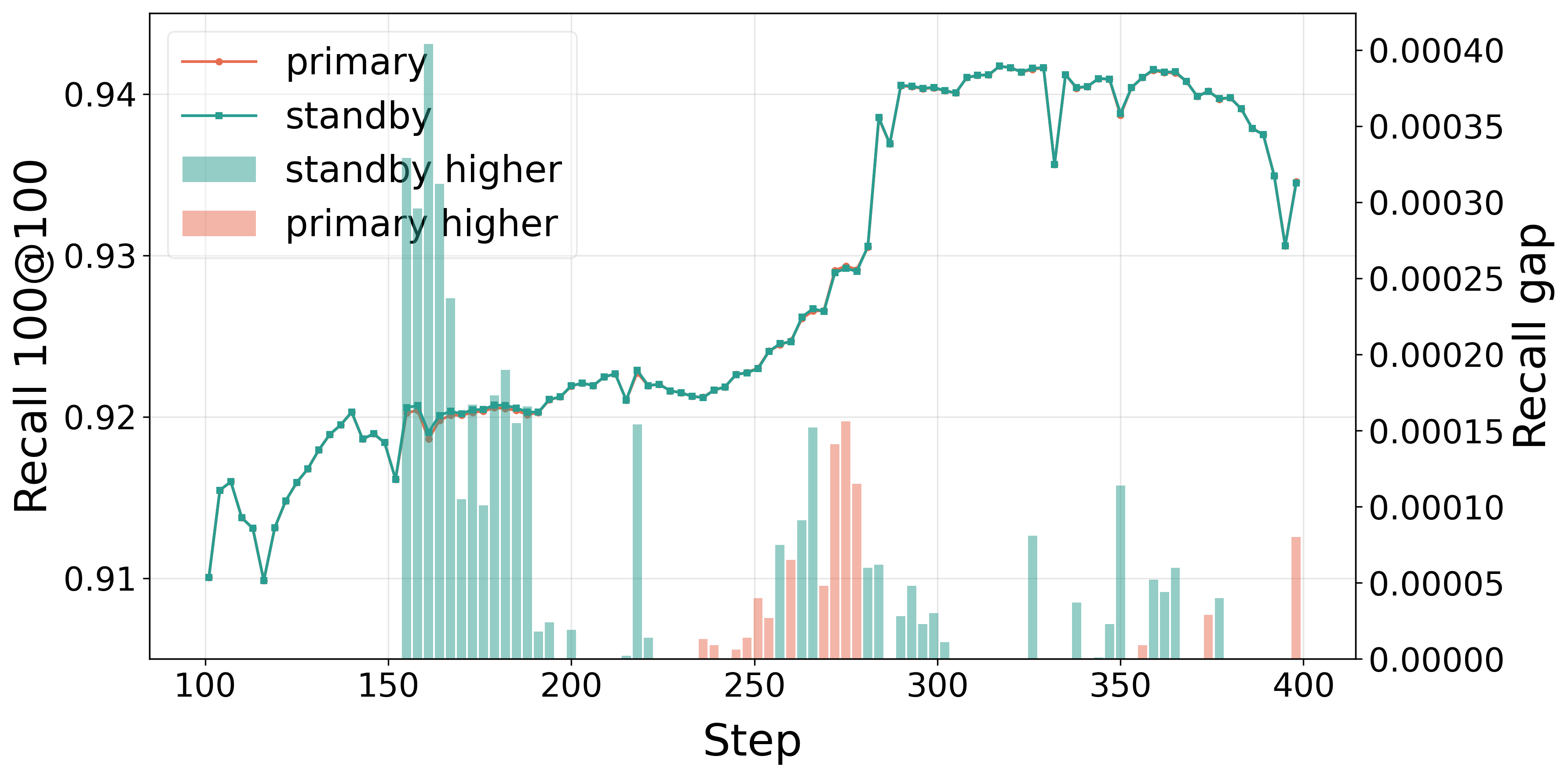}
    \caption{Recall on the Primary vs. the Standby}
    \label{fig:ev_repl-recall}
\end{figure}

We evaluate physical replication along two axes: the overhead that serving standbys imposes on the primary, and whether the standby answers queries as accurately as the primary. We deploy one primary and three standbys, driving both with the same sliding-window runbook used in Section~\ref{sec:ev_updates} at a concurrency of 32.

We first measure the overhead that serving the three standbys imposes on the primary, executing the runbook in mixed mode both with and without the standbys attached. Serving the standbys leaves throughput essentially unchanged: the primary sustains about \textbf{12.9 K ops/s} in both cases, whether running on its own or streaming WAL to all three standbys.

We next evaluate the standby's search accuracy relative to the primary. For this comparison we run the runbook in non-mixed mode, where recall is well defined, and before issuing queries on the standby we wait for it to catch up to the primary, so the two nodes are compared at the same logical point in the workload rather than across replication lag. Figure~\ref{fig:ev_repl-recall} reports the result on twin axes: the left axis carries two line series tracking Recall@100 on the primary and standby, and the right axis carries bars giving the absolute recall gap between them at each step, colored to distinguish the steps where the primary leads from those where the standby leads.

The two line series are nearly indistinguishable. Through the early phase recall is identical on both nodes, around \textbf{0.91--0.92}; once the curves begin to diverge the gap stays negligible, with a maximum of \textbf{0.0004} at any single step and a mean of about \textbf{0.00005} across the run. The standby is never systematically worse than the primary, so replication introduces no degradation in search quality, and both nodes trace the same recall trend as the standalone dynamic search results (Section~\ref{sec:ev_updates}).

A small recall gap remains even with non-mixed mode and standby catch-up, because "catch-up" only means the standby's WAL replay has reached the primary's logical point, not that it holds the same segments: segments propagate over a separate, asynchronous side channel with no timing guarantee relative to WAL replay (Section~\ref{sec:repl-design}).  A standby that has not yet adopted a freshly flushed segment instead searches the older segments or memtables, causing a recall difference. The effect is small and bounded, so reads can be offloaded to the replica without sacrificing search accuracy.

\section{Conclusion}\label{sec:conclusion}
This paper presents \sys{}, an efficient, scalable, crash-resilient integrated vector database inside PostgreSQL. Building on the decoupled index architecture of our prior \cidr{} prototype~\cite{PostgreSQLV-CIDR26}, \sys{} delivers three advances that close critical limitations in the prototype: fully concurrent vector search and updates across multiple clients, fast restart and recovery after a crash, and physical replication support that scales query throughput across standbys.

This work demonstrates that the decoupling principle extends beyond an initial proof of concept and, with carefully crafted designs, can be realized as a fully capable system within a mature relational engine. We hope that \sys{} serves as a reference point for future integrated vector database systems, demonstrating that high-performance, scalable, and highly available vector search can be achieved without sacrificing the transactional guarantees and operational simplicity of a relational engine.

\newpage

\bibliographystyle{ACM-Reference-Format}
\bibliography{paper}

@inproceedings{Milvus21,
  author       = {Jianguo Wang and
               Xiaomeng Yi and
               Rentong Guo and
               Hai Jin and
               Peng Xu and
               Shengjun Li and
               Xiangyu Wang and
               Xiangzhou Guo and
               Chengming Li and
               Xiaohai Xu and
               Kun Yu and
               Yuxing Yuan and
               Yinghao Zou and
               Jiquan Long and
               Yudong Cai and
               Zhenxiang Li and
               Zhifeng Zhang and
               Yihua Mo and
               Jun Gu and
               Ruiyi Jiang and
               Yi Wei and
               Charles Xie},
  title     = {{Milvus: A Purpose-Built Vector Data Management System}},
  booktitle = {SIGMOD},
  pages        = {2614--2627},
  year         = {2021},
}

@article{SingleStoreV24,
author = {Cheng Chen and 
          Chenzhe Jin and 
          Yunan Zhang and
          Sasha Podolsky and
          Chun Wu and
          Szu-Po Wang and
          Eric Hanson and
          Zhou Sun and
          Robert Walzer and
          Jianguo Wang
         },
title = {{SingleStore-V: An Integrated Vector Database System in SingleStore}},
year = {2024},
volume = {17},
number = {12},
pages = {3772--3785},
journal = {PVLDB},
}

@misc{Cohere,
    title={{Cohere (\url{https://huggingface.co/datasets/Cohere/beir-embed-english-v3})}}
}

@misc{DagstuhlVectorSeminar26,
	title = {{Dagstuhl Seminar 26161: Managing Vector Data for Retrieval Augmented Generation: Systems and Algorithms (\url{https://www.dagstuhl.de/26161})}},
}

@misc{BigANN,
	title = {{Billion-Scale Approximate Nearest Neighbor Search Challenge (\url{https://big-ann-benchmarks.com})}},
}

@article{VecDBPanel24,
author = {Jianguo Wang and 
          Shasank Chavan and
          Guoliang Li and
          Yannis Papakonstantinou and
          Charles Xie
         },
title = {{Vector Databases: What's Really New and What's Next?}},
year = {2024},
volume = {17},
number = {12},
pages    = {4505--4506},
journal = {PVLDB},
}

@inproceedings{VecDBRDBMSICDE24,
    author = {Yunan Zhang and Shige Liu and Jianguo Wang},
	title = {{Are There Fundamental Limitations in Supporting Vector Data Management in Relational Databases? A Case Study of PostgreSQL}},
    booktitle    = {ICDE},
    year = {2024},
    pages = {2835--2849},
}

@inproceedings{VecDBTutorial24,
  author       = {James Pan and
                  Jianguo Wang and
				  Guoliang Li},
  title        = {{Vector Database Management Techniques and Systems}},
  booktitle = {SIGMOD},
  year         = {2024},
  pages = {597--604},
}

@article{RNSGAli19,
  author    = {Cong Fu and
               Chao Xiang and
               Changxu Wang and
               Deng Cai},
  title     = {{Fast Approximate Nearest Neighbor Search With The Navigating Spreading-out
               Graph}},
  journal   = {Proceedings of the VLDB Endowment (PVLDB)},
  volume    = {12},
  number    = {5},
  pages     = {461--474},
  year      = {2019},
}

@article{HNSW18,
  author    = {Yury A. Malkov and
               Dmitry A. Yashunin},
  title     = {{Efficient and Robust Approximate Nearest Neighbor Search Using Hierarchical
               Navigable Small World Graphs}},
  journal   = {IEEE Transactions on Pattern Analysis and Machine Intelligence (TPAMI)},
  volume    = {42},
  number    = {4},
  pages     = {824--836},
  year      = {2018},
}

@misc{NVIDIAcuVS,
	title = {{NVIDIA cuVS (\url{https://developer.nvidia.com/cuvs})}},
}

@inproceedings{CAGRA2024,
  author       = {Hiroyuki Ootomo and
                  Akira Naruse and
                  Corey Nolet and
                  Ray Wang and
                  Tamas Feher and
                  Yong Wang},
  title        = {{CAGRA: Highly Parallel Graph Construction and Approximate Nearest
                  Neighbor Search for GPUs}},
  booktitle    = {ICDE},
  pages        = {4236--4247},
  year         = {2024},
}

@misc{AlloyDBVec,
	title = {{AlloyDB AI (\url{https://cloud.google.com/alloydb/ai})}},
}

@misc{GithubFaiss,
	title = {{Facebook Faiss (\url{https://github.com/facebookresearch/faiss})}},
}

@inproceedings{PASE20,
  author    = {Wen Yang and
               Tao Li and
               Gai Fang and
               Hong Wei},
  title     = {{PASE: PostgreSQL Ultra-High-Dimensional Approximate Nearest Neighbor
               Search Extension}},
  booktitle = {SIGMOD},
  pages     = {2241--2253},
  year      = {2020},
}

@inproceedings{YandexL16,
  author    = {Artem Babenko and
               Victor S. Lempitsky},
  title     = {{Efficient Indexing of Billion-Scale Datasets of Deep Descriptors}},
  booktitle = {CVPR},
  pages     = {2055--2063},
  year      = {2016},
}

@misc{SIFTData,
    title = {{SIFTData (\url{http://corpus-texmex.irisa.fr/})}}
}

@article{PanWL24,
  author       = {James Pan and
                  Jianguo Wang and
                  Guoliang Li},
  title        = {{Survey of Vector Database Management Systems}},
  journal      = {{VLDB Journal (VLDBJ)}},
  volume       = {33},
  number       = {5},
  pages        = {1591--1615},
  year         = {2024},
}

@article{StonebrakerP24,
  author       = {Michael Stonebraker and
                  Andrew Pavlo},
  title        = {{What Goes Around Comes Around... And Around..}},
  journal      = {{SIGMOD} Record},
  volume       = {53},
  number       = {2},
  pages        = {21--37},
  year         = {2024},
}

@misc{Pgvector,
	title = {{pgvector (\url{https://github.com/pgvector/pgvector})}},
}

@misc{pgvectorscale,
	title = {{pgvectorscale (\url{https://github.com/timescale/pgvectorscale})}},
}

@misc{Pinecone,
	title = {{Pinecone (\url{https://www.pinecone.io/})}},
}

@inproceedings{DiskANN19,
  author    = {Suhas Jayaram Subramanya and
               Fnu Devvrit and
               Harsha Vardhan Simhadri and
               Ravishankar Krishnawamy and
               Rohan Kadekodi},
  title     = {{Rand-NSG: Fast Accurate Billion-point Nearest Neighbor Search on a
               Single Node}},
  booktitle = {Annual Conference on Neural Information Processing Systems (NeurIPS)},
  pages     = {13748--13758},
  year      = {2019},
}

@article{JegouDS11,
  author    = {Herv{\'{e}} J{\'{e}}gou and
               Matthijs Douze and
               Cordelia Schmid},
  title     = {{Product Quantization for Nearest Neighbor Search}},
  journal   = {IEEE Transactions on Pattern Analysis and Machine Intelligence (TPAMI)},
  volume    = {33},
  number    = {1},
  pages     = {117--128},
  year      = {2011},
}

@misc{DiskANNGithub,
	title = {{DiskANN (\url{https://github.com/microsoft/DiskANN})}},
}

@inproceedings{SPFresh23,
  author       = {Yuming Xu and
                  Hengyu Liang and
                  Jin Li and
                  Shuotao Xu and
                  Qi Chen and
                  Qianxi Zhang and
                  Cheng Li and
                  Ziyue Yang and
                  Fan Yang and
                  Yuqing Yang and
                  Peng Cheng and
                  Mao Yang},
  title        = {{SPFresh: Incremental In-Place Update for Billion-Scale Vector Search}},
  booktitle    = {Proceedings of the Symposium on Operating Systems Principles (SOSP)},
  pages        = {545--561},
  year         = {2023},
}

@article{FreshDiskANN,
  author       = {Aditi Singh and
                  Suhas Jayaram Subramanya and
                  Ravishankar Krishnaswamy and
                  Harsha Vardhan Simhadri},
  title        = {{FreshDiskANN: A Fast and Accurate Graph-Based {ANN} Index for Streaming Similarity Search}},
  journal      = {CoRR},
  volume       = {abs/2105.09613},
  year         = {2021},
}

@article{LuoC20,
  author       = {Chen Luo and
                  Michael J. Carey},
  title        = {{LSM-based Storage Techniques: A Survey}},
  journal      = {{VLDB} Journal},
  volume       = {29},
  number       = {1},
  pages        = {393--418},
  year         = {2020},
}

@article{simhadri2024resultsbigannneurips23,
  author       = {Harsha Vardhan Simhadri and
                  Martin Aum{\"{u}}ller and
                  Amir Ingber and
                  Matthijs Douze and
                  George Williams and
                  Magdalen Dobson Manohar and
                  Dmitry Baranchuk and
                  Edo Liberty and
                  Frank Liu and
                  Benjamin Landrum and
                  Mazin Karjikar and
                  Laxman Dhulipala and
                  Meng Chen and
                  Yue Chen and
                  Rui Ma and
                  Kai Zhang and
                  Yuzheng Cai and
                  Jiayang Shi and
                  Yizhuo Chen and
                  Weiguo Zheng and
                  Zihao Wan and
                  Jie Yin and
                  Ben Huang},
  title        = {{Results of the Big {ANN:} NeurIPS'23 Competition}},
  journal      = {CoRR},
  volume       = {abs/2409.17424},
  year         = {2024},
}

@misc{neurips2023streaming,
    title={{NeurIPS 2023 Streaming Challenge and Beyond. (\url{https://github.com/harsha-
simhadri/big-ann-benchmarks/tree/main/neurips23/streaming.})}}
}

@misc{hnswlib,
	title = {{HNSWlib (\url{https://github.com/nmslib/hnswlib})}},
}

@misc{knowhere,
	title = {{Knowhere (\url{https://github.com/zilliztech/knowhere})}},
}

@misc{pggist, 
    title = {{GiST (\url{https://www.postgresql.org/docs/8.1/gist.html})}}, 
}

@book{Bruch24Book,
  author       = {Sebastian Bruch},
  title        = {{Foundations of Vector Retrieval}},
  publisher    = {Springer},
  year         = {2024},
}

@inproceedings{PostgreSQLV-CIDR26,
author={Jiayi Liu and Yunan Zhang and Chenzhe Jin and Aditya Gupta and Shige Liu and Jianguo Wang},
title = {{Fast Vector Search in PostgreSQL: A Decoupled Approach}},
booktitle={Conference on Innovative Data Systems Research (CIDR)},
year={2026},
}

@article{zhang2025cleannefficientdynamismgraphbased,
  author       = {Ziyu Zhang and
                  Yuanhao Wei and
                  Joshua Engels and
                  Julian Shun},
  title        = {{CleANN: Efficient Full Dynamism in Graph-based Approximate Nearest
                  Neighbor Search}},
  journal      = {CoRR},
  volume       = {abs/2507.19802},
  year         = {2025},
}

@article{Gaussdb-Vector,
author = {Sun, Ji and Li, Guoliang and Pan, James and Wang, Jiang and Xie, Yongqing and Liu, Ruicheng and Nie, Wen},
title = {GaussDB-Vector: A Large-Scale Persistent Real-Time Vector Database for LLM Applications},
year = {2025},
issue_date = {August 2025},
publisher = {VLDB Endowment},
volume = {18},
number = {12},
issn = {2150-8097},
url = {https://doi.org/10.14778/3750601.3750619},
doi = {10.14778/3750601.3750619},
journal = {Proc. VLDB Endow.},
month = aug,
pages = {4951–4963},
numpages = {13}
}

@inproceedings{MMAP_CIDR,
  title={Are you sure you want to use mmap in your database management system?},
  author={Crotty, Andrew and Leis, Viktor and Pavlo, Andrew},
  booktitle={CIDR},
  year={2022}
}

@article{gao2023retrieval,
  title={Retrieval-augmented generation for large language models: A survey},
  author={Gao, Yunfan and Xiong, Yun and Gao, Xinyu and Jia, Kangxiang and Pan, Jinliu and Bi, Yuxi and Dai, Yi and Sun, Jiawei and Wang, Meng and Wang, Haofen},
  journal={arXiv preprint arXiv:2312.10997},
  year={2023}
}

@inproceedings{karpukhin2020dense,
  title={Dense passage retrieval for open-domain question answering},
  author={Karpukhin, Vladimir and Oguz, Barlas and Min, Sewon and Lewis, Patrick and Wu, Ledell and Edunov, Sergey and Chen, Danqi and Yih, Wen-tau},
  booktitle={Proceedings of the 2020 conference on empirical methods in natural language processing (EMNLP)},
  pages={6769--6781},
  year={2020}
}

@inproceedings{covington2016deep,
  title={Deep neural networks for youtube recommendations},
  author={Covington, Paul and Adams, Jay and Sargin, Emre},
  booktitle={Proceedings of the 10th ACM conference on recommender systems},
  pages={191--198},
  year={2016}
}

@article{LSMTree,
  title={The log-structured merge-tree (LSM-tree)},
  author={O’Neil, Patrick and Cheng, Edward and Gawlick, Dieter and O’Neil, Elizabeth},
  journal={Acta informatica},
  volume={33},
  number={4},
  pages={351--385},
  year={1996},
  publisher={Springer}
}

@article{jin2026efficient,
  title={Efficient Vector Index Merging in Vector Databases},
  author={Jin, Chenzhe and Zhang, Yunan and Liu, Jiayi and Wang, Jianguo},
  journal={Proceedings of the ACM on Management of Data},
  volume={4},
  number={1 (SIGMOD},
  pages={1--26},
  year={2026},
  publisher={ACM New York, NY, USA}
}

\end{document}